\documentclass[a4paper,11pt]{article}
\usepackage{jheppub}
\usepackage{amsfonts}
\usepackage{mathtools}
\usepackage{physics}
\usepackage{array}
\usepackage{amsmath}
\usepackage{multirow}
\usepackage{caption}
\usepackage{subcaption}
\usepackage{placeins}

\usepackage{lineno}
\usepackage{adjustbox}
\usepackage{color}
\usepackage{slashed}
\usepackage{appendix}
\usepackage{bbold}
\usepackage{csquotes}
\usepackage{soul}
\usepackage[dvipsnames]{xcolor}
\usepackage{tikz}
\usepackage{float}
\usepackage[normalem]{ulem}
\usetikzlibrary{positioning,arrows}
\usetikzlibrary{decorations.pathmorphing}
\usetikzlibrary{decorations.markings}

\newcommand{\q}{q_{\rm HC}}
\newcommand{\qb}{\bar{q}_{\rm HC}}
\newcommand{\qL}{q^{L}_{\rm HC}}
\newcommand{\qbL}{\bar{q}^{L}_{\rm HC}}
\newcommand{\qbR}{\bar{q}^{R}_{\rm HC}}
\newcommand{\qR}{q^{R}_{\rm HC}}
\newcommand{\rhoT}{\rho_{\rm HC}}
\newcommand{\omT}{\omega_{\rm HC}}
\newcommand{\etaT}{\eta^{\prime}_{\rm HC}}
\newcommand{\PiT}{\Pi_{\rm HC}}
\newcommand{\FPi}{F_{\Pi}}
\newcommand{\MrhoT}{M_{\rho_{\rm HC}}}
\newcommand{\MomT}{M_{\omega_{\rm HC}}}

\newcommand{\gr}{g_{\rho_{\rm HC}}}
\newcommand{\SU}[1]{\mathrm{SU}(#1)}
\newcommand{\UN}[1]{\mathrm{U}(#1)}
\newcommand{\Sp}{\mathrm{Sp}}
\renewcommand{\tr}{\mathrm{Tr}}
\newcommand{\NHC}{N_{\rm HC}}

\newcommand{\Nf}{N_f}
\newcommand{\LTC}{\Lambda_{\rm HC}}
\newcommand{\eps}{\varepsilon}
\newcommand{\half}{\tfrac{1}{2}}

\newcommand{\LambdaT}{\Lambda_{\rm HC}}

\begin{document}

\title{\boldmath  Composite spectroscopy of a chiral $SU(4)$ model
}

\author[a]{G.~Cacciapaglia,}
\author[b,c]{A.~Deandrea,}
\author[d]{A.~M.~Iyer,}
\author[e]{S.~Kulkarni,}
\author[d]{A.~K.~Singh}
\affiliation[a]{LPTHE, UMR~7589, Sorbonne Universit\'e \& CNRS,
  4~place Jussieu, 75252~Paris Cedex~05, France; and
  QTC \& D-IAS, Southern Denmark University, Campusvej~55,
  5230~Odense~M, Denmark}
\affiliation[b]{Universit\'e Lyon 1, CNRS, IP2I, UMR 5822, Villeurbanne, France}
\affiliation[c]{Department of Physics, University of Johannesburg,
PO Box 524, Auckland Park 2006, South Africa}
\affiliation[d]{Department of Physics, IIT Delhi,
  New Delhi~110016, India}
\affiliation[e]{Institute of Physics, NAWI Graz, University of Graz,
  Universit\"atsplatz~5, A-8010~Graz, Austria}

\emailAdd{cacciapa@lpthe.jussieu.fr}
\emailAdd{deandrea@ip2i.in2p3.fr}
\emailAdd{iyerabhishek@physics.iitd.ac.in}
\emailAdd{suchita.kulkarni@uni-graz.at}
\emailAdd{Abhishek.Kumar.Singh@physics.iitd.ac.in}

\abstract{
We consider a QCD-inspired chiral $SU(4)$ model, with a new fundamental sector ($q_{\rm HC}$)  charged under the confining hypercolor gauge group $SU(3)_{\rm HC}$  as a template for electroweak dynamics and compositeness. The $(\SU{4}_L\times\SU{4}_R)/\SU{4}_V$ breaking leads to a rich structure of spin-0 and spin-1 resonances which can be produced by the hadronization of $q_{\rm HC}$ in Drell-Yan processes at FCC-hh. Working in the true vacuum at a general misalignment angle $\theta$, the fermion currents are derived for the fifteen techni-rho ($\rhoT$) resonances, the techni-omega ($\omT$), and the techni-eta-prime ($\etaT$) using the Hidden Local Symmetry formalism. The estimated branching fractions pave the way for studying the model spectroscopy. The analysis of the production by hadronization of  $\rhoT$, $\omT$ and $\etaT$ states is presented for a 100 TeV future hadronic collider, FCC-hh. The projected discovery of $\rhoT$ and $\etaT$ is complementary to the direct production by Drell-Yan or Vector Boson Fusion, providing a smoking gun signature for Higgs compositeness.}

\maketitle

\section{Introduction}
\label{sec:intro}
The discovery of a Higgs boson~\cite{ATLAS:2012yve,CMS:2012qbp}, with couplings close to the predicted ones, established that the  electroweak symmetry breaking (EWSB) originates from  the vacuum expectation value of a scalar doublet~\cite{Higgs:1964pj}, whose origin, however, remains undisclosed. The Higgs mechanism of the Standard Model (SM) is phenomenological~\cite{PhysRevLett.13.321,Higgs:1964pj}, with no explicit construction of an underlying dynamics driving its breaking. One possibility is to consider electroweak (EW) compositeness modelled on quantum chromodynamics (QCD). A  new strongly coupled gauge interaction, the Hypercolor (HC) group $G_{\rm HC}$, binds new fundamental fermions (hyperquarks) whose condensate dynamically breaks the EW symmetry~\cite{Weinberg:1975gm,Weinberg:1979bn,Susskind:1978ms,Farhi:1980xs}, precisely as the QCD condensate $\langle \bar{q}q \rangle$ spontaneously breaks the chiral symmetry of light quarks. In the minimal framework with two fundamental fermions, the original historical version of Technicolor (TC)~\cite{Weinberg:1975gm,Susskind:1978ms}, the breaking of the global symmetry ($\mathcal{G}$) to its vectorial subgroup ($\mathcal{H}$),  at a scale $\Lambda_{\rm HC}\sim 4\pi v$,  also breaks the EW symmetry, and the three would-be Goldstone techni-pions are absorbed by $W^\pm$ and $Z$ bosons~\cite{Weinberg:1979bn,Susskind:1978ms,Dimopoulos:1979es,Eichten:1979ah}. Several extensions to this framework, with specific choices of $\mathcal{G}/\mathcal{H}$, were considered for generating the SM fermion masses~\cite{Dimopoulos:1979es,Eichten:1979ah,Kaplan:1991dc}, consistency with EW precision tests~\cite{Peskin:1990zt,Peskin:1991sw}, and are comprehensively reviewed in \cite{Farhi:1980xs,Hill:2002ap,Cacciapaglia:2020kgq}. In the following, we will denote as ``TC-like'' all the models in this class, where $\Lambda_{\rm HC} \sim v$ and the scalar doublet can be identified with the condensate of hyperquarks.

The composite Higgs (CH) framework~\cite{Kaplan:1983fs,Georgi:1984af,Kaplan:1983sm} is another manifestation of this idea in models where there exists a condensate that preserves the EW symmetry. Hence, the scalar doublet emerges as a pseudo-Nambu-Goldstone boson (pNGB) breaking a global symmetry $\mathcal{G}$, of the new strong sector, to a subgroup $\mathcal{H}$ at scale $\Lambda_{\rm HC}\sim 4\pi f > 4\pi v$, where $f$ is the pNGB decay constant. In these models, the CH and TC scenarios represent two different extreme cases, where the breaking of $\mathcal{G}$ preserves the EW symmetry in the former, while it is broken completely in the latter. The scale $f$ is related to the vacuum expectation value as $v = f\sin\theta$~\cite{Kaplan:1983fs}. The limits $\theta \to \pi/2$ and $\theta \to 0$ recover respectively the TC limit and the CH limit~\cite{Galloway:2010bp,Cacciapaglia:2014uja}. 

The choice of coset $\mathcal{G}/\mathcal{H}$ determines the spectrum of the pNGBs. Remaining within the realm of HC theories, there exist three coset classes, corresponding to the reality of the fermion representation under the fundamental gauge group~\cite{Ferretti:2013kya}. For a theory with $\Nf$ fermion flavours of the same species, real \cite{Dugan:1984hq,Agugliaro:2019wtf} or pseudo-real \cite{Ryttov:2008xe,Cacciapaglia:2014uja} representations realize a global flavour symmetry $\mathcal{G}=\mathrm{SU}(2\Nf)$ broken to $\mathcal{H}= \mathrm{SO}(2\Nf)$ or $\mathrm{Sp}(2\Nf)$, respectively. The complex representation gives the familiar QCD-like scenario, with flavour symmetry $\mathrm{SU}(\Nf)_L\times \mathrm{SU}(\Nf)_R$ broken to $\mathrm{SU}(\Nf)_V$~\cite{Ma:2015gra}.

In the present work, we consider the `minimal version' of theories featuring both a TC and CH limit with the complex representation: the $(\SU{4}_L \times \SU{4}_R) / \SU{4}_V$ model~\cite{Ma:2015gra,Wu:2017iji}. This scenario is the QCD-like generalization of the two-flavour chiral limit to the four-flavour case, and the breaking $\SU{4}_L \times \SU{4}_R \to \SU{4}_V$ produces $15$ pNGBs (techni-pions) $\PiT$ in the adjoint of $\SU{4}_V$, together with a spectrum of fifteen techni-rho $\rhoT$ spin-1 resonances, one techni-omega $\omT$, and one techni-eta-prime $\etaT$. Working in the true vacuum at a general misalignment angle $\theta$ and using the Hidden Local Symmetry (HLS) formalism~\cite{Bando:1985rf,Bando:1987br}, we derive the fermion currents for all fifteen $\rhoT$ resonances, the $\omT$, and the $\etaT$, and show that the branching fractions reveal a sharp selection rule: only the isotriplet $\rhoT^{\pm,0}$ couples to $W^\pm$ and $Z$ bosons, and this coupling is independent of $\theta$, making it a universal prediction present in both the TC and CH limits. 

The spin-1 resonances, i.e. the techni-rho and techni-omega,
and the pseudoscalar techni-eta-prime have a long history of collider phenomenology studies~\cite{Eichten:1997yc,Lane:1999uh,Lane:2002sm,Chivukula:1995dt,Chivukula:1989rn,Lane:2009ct,Holdom:1995hd,PhysRevD.96.075040}.
Additionally, individual searches for some of these resonances are ongoing at the LHC~\cite{ATLAS:2020fry,CMS:2026fjw,ATLAS:2024qvg,ATLAS:2018ocj}; conclusions derived from these analyses can place bounds on the masses and couplings, without delving into the details of the theory that may have led to these resonances~\cite{Pappadopulo_2014,PhysRevD.92.085019,Belyaev_2009,Baker_2022}. In fact, similar signals could emerge in other theories without compositeness, for instance based on extended gauge symmetries or extra dimensions~\cite{Langacker:2008yv,Agashe:2003zs}. The only way to probe the composite nature of the EWSB relies on probing the theory at energy scales above the confinement scale $\Lambda_{\rm HC}$~\cite{Cacciapaglia:2025multiboson}. The second part of this work develops a more detailed study using the extended possibilities offered by future multi-TeV colliders.
Using the formalism proposed in~\cite{Cacciapaglia:2025multiboson},
we consider pair production by means of the fragmentation and hadronization of hyperquarks, in the process $pp \to \q\; \bar{q}_{\rm HC}$ above  the confinement scale  $\Lambda_{\rm HC}\sim \mathcal{O}$(TeV) at a future 100 TeV collider, akin to the FCC-hh project~\cite{FCC:2018byv}.  

While identifying the final states of the entire spectrum,  we demonstrate the detailed collider reach for a fraction of them. This method is robust and easily extendable to other resonances in the spectrum. This strategy would not only be a smoking gun for EW compositeness but would also provide hints for the underlying flavour and gauge group.  

The paper is organized as follows: Section~\ref{sec:model} presents the
theoretical framework of the model, discussing the symmetry-breaking pattern,
the mass spectrum, the unbroken generators at a general vacuum misalignment
angle $\theta$, and the vector currents. Section~\ref{sec:HLS} develops the
spin-1 resonances in the Hidden Local Symmetry (HLS) formalism.
Section~\ref{sec:branching} discusses the dominant decay channels and branching
ratios of the spin-1 resonances. Section~\ref{sec:pheno} turns to collider
phenomenology: we first collect the bounds that the LHC already places on these
states, and then study the production of the spin-1 and spin-0 resonances
through hyperquark fragmentation and hadronization at the FCC-hh. We conclude in
Section~\ref{sec:conclusion}. Technical details are collected in four
appendices: the generator basis, vacuum structure and commutation relations in
Appendix~\ref{app:generators}, the HLS derivations in Appendix~\ref{app:HLS}, the group weights and the branching-ratio formula in Appendix~\ref{app:phasespace}, and kinematics in Appendix~\ref{app:kinematics}.\\

\section{Theoretical Framework}
\label{sec:model}
\subsection{A chiral $\mathrm{SU}(4)$ model with four Dirac hyperquarks}
\label{sec:model:coset}

The formation of bound states from quarks in the fundamental representation of $\mathrm{SU}(3)_C$ is well modelled in QCD.
Following the same logic, we consider a QCD-like theory in which the underlying dynamics is an asymptotically free $\mathrm{SU}(N_{\rm HC})$ hypercolor gauge theory with $N_f = 4$ Dirac hyperquarks in the fundamental representation~\cite{Ma:2015gra,Wu:2017iji}.
In the chiral limit the hypercolor Lagrangian is
\begin{equation}
 \mathcal L_{\rm HC} = -\tfrac14 G^a_{\mu\nu}G^{a\mu\nu}
  +\sum_{i=1,2}\bigl(\bar U_i\,i\slashed D\,U_i + \bar D_i\,i\slashed D\,D_i\bigr),
\end{equation}
where $D_\mu = \partial_\mu - ig_{\rm HC}\,G^a_\mu\,T^a_{\rm HC}$.
The global symmetry of this Lagrangian is
\begin{equation}
\mathbf{G}= \mathrm{SU}(4)_L \times \mathrm{SU}(4)_R \times \mathrm{U}(1)_A \times \mathrm{U}(1)_V, \label{eq:Gflavour}
\end{equation}
as can be seen by combining the left-handed and right-handed components of the Dirac fermions in 4-component fields, $\q^{L/R} = (U_1, D_1, U_2, D_2)_{L/R}$, where $\qL$ transforms as the $\mathbf{4}$ of $\mathrm{SU}(4)_L$ and $\qR$ as the $\bar{\mathbf{4}}$ of $\mathrm{SU}(4)_R$.
The strong hypercolor dynamics induces a bilinear condensate near the confinement scale $\Lambda_{\rm HC}$.
Organizing the four flavours into a single column vector $\q = (\qL, \qR)^T$, the hyperquark condensate can be written as an $8\times 8$ matrix 
\begin{equation}
  \langle \qb\, \q \rangle = \langle \bar{q} q\rangle\  \Sigma_0\,, \quad \mbox{where}\;\; \Sigma_0 \equiv
  \begin{pmatrix}
      0 & \Sigma^4_0 \\
      \Sigma^4_0 & 0
  \end{pmatrix},   
  \label{eq:condensate}
\end{equation}
and the flavour-traced condensate $\langle \bar{q} q \rangle \equiv - \FPi^2 \LTC$ sets the physical scales ($\FPi$ being the decay constant of the pNGBs).
Here $\Sigma_0$ is a unitary $8\times 8$ matrix in the combined left-right flavour space, whose off-diagonal block structure reflects that only Lorentz-invariant pairings $\qbL \qR$ and $\qbR \qL$ acquire a vacuum expectation value. The precise form of the $4\times 4$ matrix $\Sigma_0^4$ depends on the alignment of the vacuum with respect to the EW charges of the fermions, while the breaking of the chiral symmetry is independent of such details. Hence, the condensate breaks the global flavour symmetry
$\mathcal{G} \equiv \mathrm{SU}(4)_L \times \mathrm{SU}(4)_R \rightarrow$ 
$\mathcal{H} \equiv \mathrm{SU}(4)_V$, giving fifteen Goldstone bosons, each
associated with a broken generator of the coset $\mathcal{G}/\mathcal{H}$.
In addition, the spectrum of vector resonances consists of fifteen states
associated with the unbroken generators of the subgroup $\mathcal{H}$, and a singlet
$\omega_{\rm HC}$ associated with $\mathrm{U}(1)_V$.
The isosinglet $\eta'_{\rm HC}$, associated with the broken
$\mathrm{U}(1)_A$, is not a true Goldstone boson as $\mathrm{U}(1)_A$ is
broken explicitly through the hypercolor anomaly~\cite{Witten:1979vv,Veneziano:1979ec},
acquiring a mass of order $\Lambda_{\rm HC}$.

\subsection{Mass spectrum of the spin-0 and spin-1 resonances}
\label{sec:model:spectrum}

The model spans a family of scenarios parametrized by the misalignment angle $\theta$,
interpolating between the EW-preserving vacuum at $\theta = 0$
and the TC limit at $\theta = \pi/2$ (where $\sin\theta =1$).
The pNGB decay constant is related to the Fermi scale $v$ by
\begin{equation}
  \sqrt{2}\,\FPi\,\sin\theta = v \approx 246\,\mathrm{GeV},
  \qquad
  \FPi\Big|_{\rm TC} = \frac{v}{\sqrt{2}} \approx 174\,\mathrm{GeV}.
  \label{eq:TCfpi}
\end{equation}
Independently of the alignment of the vacuum, as long as $\theta \neq 0$, three Goldstones transforming as an EW isotriplet are absorbed by the $W^\pm$ and $Z$ as their longitudinal polarizations.
The remaining twelve heavy techni-pions remain as physical states and acquire masses near the EW scale. By scaling the QCD pion masses, we can estimate 
\begin{equation}
M_{\PiT} \sim m_\pi\,\left(\frac{\FPi}{f_\pi}\right) \approx  \frac{250\,\mathrm{GeV}}{\sin\theta},
\label{eq:mass_pi}
\end{equation}
where $m_\pi = 135$~MeV and $f_\pi = 93$~MeV, so they sit well below the confinement scale $\LTC$.
All spin-1 resonances, by contrast, are bound states of the hypercolor dynamics and
therefore live at the confinement scale $\LTC$; their masses are obtained in the large $N$ limit by scaling from
QCD. Following~\cite{Farhi:1980xs,RevModPhys.56.579,Hill:2002ap} we obtain
\begin{equation}
  \MrhoT = \MomT
  \;\simeq\; m_\rho\,\frac{\FPi}{f_\pi}\;\sqrt{\frac{3}{\NHC}}
  \;\sim\; \frac{2\,\mathrm{TeV}}{\sin\theta},
  \label{eq:mass_rho}
\end{equation}
where $m_\rho = 770\,\mathrm{MeV}$, $\NHC = 3$, and $f_\pi = 93\,\mathrm{MeV}$ are the
QCD rho mass, number of hypercolors, and pion decay constant, respectively. 
The $\rhoT$ and $\omT$ are degenerate at leading order in the large-$\NHC$ expansion;
their mass splitting is a sub-leading $1/\NHC$ effect.
The spectrum therefore has a clear two-scale structure: in the TC limit, heavy techni-pions near $250\,\mathrm{GeV}$
and all vector resonances near $\LTC \approx 2\,\mathrm{TeV}$, with
$\MrhoT \simeq \MomT \sim 2\,\mathrm{TeV}$, which is the scale
targeted by the FCC-hh analysis of Section~\ref{sec:pheno}. In the CH limit, all scales are lifted by a common factor $1/\sin\theta$, where $\sin\theta \gtrsim 0.1$: this allows us to set a second benchmark with $\MrhoT \simeq \MomT \sim 5\,\mathrm{TeV}$, cf. Section~\ref{sec:pheno}.

The $\etaT$ is not a true Goldstone boson, since $\UN{1}_A$ is broken explicitly by the
hypercolor anomaly. It acquires a mass through the Witten--Veneziano
mechanism~\cite{Witten:1979vv,Veneziano:1979ec} (see also~\cite{DiVecchia:2013swa}),
\begin{equation}
  M^2_{\etaT}
  = \frac{2\Nf}{\FPi^2}\,\chi_{\rm HC}
  + M^2_{\rm ETC},
  \label{eq:meta_mass}
\end{equation}
where $\chi_{\rm HC} \sim \LTC^4$ is the topological susceptibility of the pure
hypercolor Yang--Mills theory~\cite{Ce:2016awn,DelDebbio:2004ns,Lucini:2012gg} and $M^2_{\rm ETC}$ accounts for the explicit breaking needed for the SM fermion mass generation, such as ETC interactions. Since $\chi_{\rm HC}$ is independent of 
$\NHC$ at leading order while $\FPi^2 \propto \NHC$,
the $\etaT$ mass-squared is parametrically of order $1/\NHC$, making it lighter than
the $\rhoT$ and $\omT$ at large $\NHC$, in direct analogy with the suppression of the
QCD $\eta'$ mass in the large $N$ limit~\cite{DIVECCHIA1980253,DIVECCHIA1980247}. Rescaling the QCD value accordingly~\cite{Tandean:1995ci,PhysRevD.96.075040},
\begin{equation}
  M_{\etaT} \approx \frac{\FPi}{f_\pi}\,\frac{\sqrt{3\Nf}}{\NHC}\,m_{\eta^{0}} \approx \frac{2\,\mathrm{TeV}}{\sin\theta},
  \label{eq:meta_scaling}
\end{equation}
where $\NHC = 3$, $\Nf = 4$ is the number of hyperquark flavours, and $m_{\eta^{0}} \simeq 849$~MeV is the mass of the flavour-singlet pseudoscalar in the chiral limit of $\SU{3}$-flavour QCD.

For completeness, the quark condensate can be estimated in analogy to the current algebra Gell-Mann, Oakes, and Renner (GMOR) relation in QCD~\cite{Gell-Mann:1968hlm,Manohar:1983md}, yielding \begin{equation}
\langle\bar{q}q\rangle \equiv -\left. \frac{M_{\PiT}^2 \FPi^2}{N_f\ m_{\q}} \right|_{m_{\q} \to 0} \sim -\FPi^2 \Lambda_{\rm HC},
\end{equation}
where $m_{\q}$ is the Dirac mass of the techniquarks.

\subsection{Vacuum alignment, unbroken generators, and vector currents}
\label{sec:model:breaking}

In order to properly describe the spin-1 resonances and their associated currents, we need to define a specific vacuum alignment in the flavour space of Eq.~\eqref{eq:Gflavour}. The first step is to define the EW charges of the hyperquarks $U_{1,2}$ and $D_{1,2}$: in our convention, the two Dirac fermions $U_1$ and $D_1$ form a doublet of $\SU{2}_L$, while $U_2$ and $D_2$ form a doublet of $\SU{2}_R$ (i.e., two singlets with hypercharges $\pm1/2$). The most general vacuum is characterized by an angle $\theta$ with respect to the alignment that preserves the EW symmetry~\cite{Ma:2015gra}.
The true vacuum in the $\SU{4}$ space is given by
\begin{equation}
  \Sigma^{4}_{0} (\theta) = \Omega^{2}_{4},
  \qquad
  \Omega_4 \;=\;
  \begin{pmatrix}
    \cos\tfrac{\theta}{2}\,I_2 & \sin\tfrac{\theta}{2}\,I_2 \\[4pt]
    -\sin\tfrac{\theta}{2}\,I_2 & \cos\tfrac{\theta}{2}\,I_2
  \end{pmatrix},
\end{equation}
which can be plugged into the $8\times 8$ block matrix $\Sigma_0$~\cite{Preskill:1980mz} defined in Eq.~\eqref{eq:condensate}. Note that for any value $0 < \theta \leq \pi/2$, this vacuum would break the custodial symmetry of the SM in the same way as the Higgs does, i.e. $\SU{2}_L\times\SU{2}_R \to \SU{2}_V$.
At $\theta = 0$ the EW symmetry is unbroken, while $\theta = \pi/2$ corresponds to the TC limit where $\sqrt{2}\,\FPi$ plays the role of the Higgs vacuum expectation value $v$. In the intermediate case $0\leq \theta \leq \pi/2$, we have $\sqrt{2}\,\FPi\ \sin\theta = v$. 
The fifteen unbroken generators $\widetilde{\mathcal{T}}^{a}$ in the misaligned vacuum satisfy
\begin{equation}
  \widetilde{\mathcal{T}}^{a}\Sigma_{0}
  + \Sigma_{0}\widetilde{\mathcal{T}}^{a\,T} = 0,
\end{equation}
and their $8\times8$ matrix representation is
\begin{equation}
  \widetilde{\mathcal{T}}^{a} =
  \begin{pmatrix}
    \widetilde{\Lambda}^{a} & 0 \\
    0 & -(\widetilde{\Lambda}^{a})^{T}
  \end{pmatrix},
  \label{eqs:generator}
\end{equation}
where $\widetilde{\Lambda}^{a}$ are the $\mathrm{SU}(4)$ generators in the rotated basis of $\Omega_4$.
All fifteen explicit generator matrices are given in Appendix~\ref{app:generators}.

The spin-1 resonances transform as the adjoint of $\mathrm{SU}(4)_V$ and decompose under the custodial symmetry as follows, with quantum numbers labelled as $[\SU{2}_V]_{(\SU{2}_L,\SU{2}_R)}$:
\begin{equation}
  \mathbf{15}_{\mathrm{SU}(4)_V}
  = [\mathbf{3}]_{(3,1)}
    \oplus [\mathbf{3}]_{(1,3)}
    \oplus ([\mathbf{3}]\oplus[\mathbf{1}])_{(2,2)_I}
    \oplus ([\mathbf{3}]\oplus[\mathbf{1}])_{(2,2)_{II}}
    \oplus [\mathbf{1}]_{(1,1)}.
\end{equation}
Using the first six generators $\widetilde{\mathcal{T}}^{a=1,\ldots,6}$, we form the vector $T^{V,(a)} = \widetilde{\mathcal{T}}^a + \widetilde{\mathcal{T}}^{a+3}$ and axial $T^{A,(a)} = \widetilde{\mathcal{T}}^a - \widetilde{\mathcal{T}}^{a+3}$ linear combinations (for $a=1,2,3$) that close under the Lie algebra of $\mathrm{SU}(2)_V$, together with the remaining generators as defined in Table~\ref{tab:commutators_app} in Appendix~\ref{app:generators}.
Based on the definition of the misaligned unbroken generators in Eq.~\eqref{eqs:generator}, the vector current for each resonance is given by
\begin{equation}
  \rho^{\mu}{}^{a}
  = \qbL\,\widetilde{\Lambda}^a\,\gamma^\mu\,\qL
  - \qbR\,(\widetilde{\Lambda}^a)^T\,\gamma^\mu\,\qR.
  \label{eq:current_formula}
\end{equation}
Using this current formula together with the linear combinations of the generators and their ladder operators, we compute all fifteen currents of the spin-1 resonances, collected in Table~\ref{tab:currents_app} of Appendix~\ref{app:generators}.
The current of $\omega_{\rm HC}$, associated with $\mathrm{U}(1)_V$, is
\begin{equation}
  \omega^{\mu} = \frac{1}{2\sqrt{2}}\Bigl[
    \bar{U}_1\gamma^\mu U_1
  + \bar{D}_1\gamma^\mu D_1
  + \bar{U}_2\gamma^\mu U_2
  + \bar{D}_2\gamma^\mu D_2
  \Bigr],
\end{equation}
and the axial current of the $\eta'_{\rm HC}$, the pNGB of the broken $\mathrm{U}(1)_A$, is
\begin{equation}
  \eta^{\mu}
  = \frac{1}{2\sqrt{2}}\Bigl[
    \bar{U}_1\gamma^\mu\gamma^5 U_1
  + \bar{D}_1\gamma^\mu\gamma^5 D_1
  + \bar{U}_2\gamma^\mu\gamma^5 U_2
  + \bar{D}_2\gamma^\mu\gamma^5 D_2
  \Bigr].
\end{equation}
The three currents associated with the generators $T^{V,(a)}$,
that is $\rhoT^+$, $\rhoT^-$, and $\rhoT^0$, are
\begin{equation}
\begin{aligned}
  \rho^{\mu,+} &= \frac{1}{2}\Bigl[
    \bar{U}_1\gamma^\mu D_1 + \bar{U}_2\gamma^\mu D_2
  \Bigr],
  \\[2pt]
  \rho^{\mu,-} &= \frac{1}{2}\Bigl[
    \bar{D}_1\gamma^\mu U_1 + \bar{D}_2\gamma^\mu U_2
  \Bigr],
  \\[2pt]
  \rho^{\mu,0} &= \frac{1}{2\sqrt{2}}\Bigl[
    \bar{U}_1\gamma^\mu U_1 + \bar{U}_2\gamma^\mu U_2
  - \bar{D}_1\gamma^\mu D_1 - \bar{D}_2\gamma^\mu D_2
  \Bigr],
\end{aligned}
\label{eq:rho_iso_currents}
\end{equation}
which are notably independent of $\theta$.
Their fermion bilinears have exactly the same structure as the QCD $\rho$ mesons: the
charged currents pair hyperquarks of different flavour within the same $\SU{2}$
doublet, while the neutral current is a flavour-diagonal vector current, and together
the three states form an isospin triplet of $\SU{2}_V$, precisely as in the QCD
case~\cite{GellMann:1962xb,Sakurai:1960ju}.
These three states share the same G-parity quantum number as the QCD $\rho$ meson, a discrete symmetry of the hypercolor dynamics that forbids its mixing with the
isosinglet $\omT$ up to EW-breaking effects~\cite{Lee:1956sw}.
The three states are, therefore, identified as the direct QCD-like hypercolor isotriplet,
and in Eq.~\eqref{eq:rho_iso_currents} and Table~\ref{tab:currents_app} they are labelled without a hat as
$\rhoT^{\pm,0}$.
Their current structure, and hence their couplings to the EW sector, is entirely
independent of the misalignment angle $\theta$, present with the same strength in both the
pure TC limit ($\theta \to \pi/2$) and the pure CH limit ($\theta \to 0$), making them a
universal prediction of the hypercolor sector~\cite{Franzosi:2016aoo}.
 
Two further sets of states in Table~\ref{tab:currents_app} are also $\theta$-independent: the triplet $\rhoT^{X^{\mathrm{II},(i)}}$ (for $i = 1,2,3$) and the singlet $\rhoT^{S^{\mathrm{II}}}$, associated with the generators $X^{\mathrm{II},(i)}$ and $S^{\mathrm{II}}$, respectively.
Their currents consist of purely imaginary antisymmetric bilinears mixing the two $\SU{2}$ doublets, and in the rotated basis they commute with the vacuum rotation $\Omega_4$, so no factors of $\theta$ enter.
These states are also labelled without a hat in Table~\ref{tab:currents_app} to signal their $\theta$-independence.
All remaining states, those associated with $T^{A,(a)}$, $X^{\mathrm{I},(i)}$, $S^{\mathrm{I}}$, and $S^{0}$, carry explicit factors of $\cos\theta$ and $\sin\theta$ in their currents, meaning that their coupling strengths interpolate between the TC and CH limits and can in principle serve as discriminators between the two scenarios.
These states are labelled with a hat in Table~\ref{tab:currents_app} to make explicit the $\theta$-dependence.
An analogous current classification was carried out for the $\SU{4}/\Sp(4)$ CH model in~\cite{Franzosi:2016aoo}, where the HLS formalism was used to derive the spin-1 currents and their couplings to the EW sector, and our
$(\SU{4}_L\times\SU{4}_R)/\SU{4}_V$ case extends that construction
to the chiral breaking pattern with fifteen resonances. 
 The $\theta$ dependence is present in only some of the states in Table~\ref{tab:currents_app}. This can be understood from the effective field theory picture of vacuum misalignment in CH models (see for example \cite{Cacciapaglia:2014uja,Franzosi:2016aoo}). Vacuum alignment affects states according to how the corresponding operators transform under the rotation that misaligns the vacuum. If a composite state/operator is invariant under the full symmetry $\mathcal{G}$, it will not be affected by a vacuum rotation, which is just a field redefinition from the point of view of this $\mathcal{G}$-invariant term. If the operator breaks $\mathcal{G}$ then it may contain $\theta$ terms when it is rotated, depending on whether it commutes or not with the particular alignment generator. Note also that different components of the same multiplet can have different $\theta$ dependence because the explicit-breaking spurions distinguish their quantum numbers. Another important remark is that a state can have a $\theta$-independent mass but still have $\theta$-dependent couplings.
 
Since the couplings of $\rhoT^{\pm,0}$ to the EW bosons do not depend on $\theta$, its branching fraction into $W^\pm Z$ and $W^+W^-$ is equally $\theta$-independent: it is determined purely by the $\SU{4}$ group factors and the $P$-wave phase space, without reference to the degree of vacuum misalignment. 
To make this quantitative, we need the effective interaction Lagrangian for the
$\rhoT$ -- $\PiT$ -- $\PiT$ vertex, which we now derive using
the HLS formalism.

\section{The Spin-1 Resonances in the Hidden Local Symmetry Formalism}
\label{sec:HLS}
The HLS formalism provides an effective field-theoretic
description of vector mesons in theories with spontaneous chiral symmetry
breaking~\cite{Bando:1985rf,BANDO1985493,Bando:1987br,Ecker:1989yg,Georgi:1989xy}. Originally
developed for QCD, where the $\rho$ mesons emerge as the gauge bosons of a hidden
$\SU{2}_V$ local symmetry, it reproduces the universality of the $\rho$
coupling, the KSRF relation~\cite{PhysRevLett.16.255,PhysRev.147.1071} between the $\rho$ mass and the pion decay constant, and the vector meson dominance
picture~\cite{Harada:2003jx}. The same formalism has subsequently been extended to CH
models, including the $\SU{4}/\Sp(4)$ coset~\cite{Franzosi:2016aoo}. In this
section we apply the HLS construction to the
$(\SU{4}_L \times \SU{4}_R) / \SU{4}_V$ hypercolor model. 

To implement the HLS formalism in our model, it is convenient to describe the fifteen techni-pions by the
non-linearly realized field $U$, expressed as a $4\times 4$ matrix as 
\begin{equation}
U = \exp\!\left(\frac{2i\Pi}{\FPi}\right)
\end{equation}
with $\Pi = \PiT^a\widetilde{\Lambda}^a$ (for $a = 1,\dots,15$) being the Nambu--Goldstone
matrix transforming as $\Pi \to g_L\, \Pi\, g_R^\dagger$, where $\widetilde{\Lambda}^a$ are the $4\times4$  generators of $\SU4$ defining $\widetilde{\mathcal{X}}^a$, as given in Appendix~\ref{app:generators}.
This non-linear description leaves a freedom in the choice of field variables. Just as in the two-flavour QCD case, where the unbroken $\SU{2}_V$ can be promoted to a local symmetry without altering the physical pion content of the theory~\cite{Bando:1985rf,Bando:1987br}, the unbroken diagonal subgroup $\SU{4}_V$ of our model can equally be gauged. This introduces a new set of vector gauge fields while leaving the techni-pion spectrum and its interactions unchanged at the level of the leading-order Lagrangian. This observation, which underlies the HLS framework, motivates combining the original global chiral symmetry
$\mathcal{G}_0=\left.\SU{4}_L\times\SU{4}_R\right|_\mathrm{global}$ with a new local copy $\mathcal{H}_1=\left.\SU{4}_V\right|_\mathrm{local}$, so that the extended theory is invariant under the enlarged symmetry group
\begin{equation}
  \mathcal{G}_0 \times \mathcal{H}_1
  \;=\;
  \left.\SU{4}_L \times \SU{4}_R\right|_\mathrm{global}
  \;\times\;
  \left.\SU{4}_V\right|_\mathrm{local}.
  \label{eq:HLS_group}
\end{equation}
The fifteen $\rhoT^a$ resonances are then identified as the dynamical gauge bosons of the hidden local symmetry $\mathcal{H}_1$, each associated with one of the unbroken generators of $\SU{4}_V$, with gauge field $\rhoT^{\mu}=\rhoT^{a\mu}\widetilde{\Lambda}^a$, where $\widetilde{\Lambda}^a$ is the $4\times4$ block of $\widetilde{\mathcal{T}}^a$ defined in Eq.~\eqref{eqs:generator}.

To realize the enlarged symmetry, we split the Nambu--Goldstone field into two chiral fields $\xi_{L,R}$, satisfying $U=\xi_L^\dagger\xi_R$ and parametrized as $\xi_{L,R} = e^{\mp i \PiT/\FPi}$, where we have fixed the unitary gauge of the hidden local symmetry. Under $\mathcal{G}_0\times \mathcal{H}_1$ they transform as
\begin{equation}
  \xi_{L,R} \;\to\; h(x)\,\xi_{L,R}\,g_{L,R}^\dagger,
  \qquad g_{L,R}\in \mathcal{G}_0,\quad h(x)\in \mathcal{H}_1,
  \label{eq:xi_transf}
\end{equation}
which reproduces the transformation of $U$, the local factor $h(x)$ cancelling in the product $\xi_L^\dagger\xi_R$. Invariance under the local factor requires the $\rhoT^{\mu}$ fields to transform inhomogeneously,
\begin{equation}
  \rhoT^\mu \;\to\; h\,\rhoT^\mu\,h^\dagger
  \;+\; \frac{i}{\gr}\,h\,\partial^\mu h^\dagger,
  \label{eq:rho_transf}
\end{equation}
which is precisely the transformation of a gauge connection of
$\mathcal{H}_{1}$ with coupling $\gr$, justifying their
identification as the spin-1 bound states of the hypercolor sector, in direct analogy with the QCD $\rho$ mesons~\cite{Bando:1985rf,Harada:2003jx}.

The chiral fields $\xi_{L,R}$ provide the building blocks for constructing the HLS-invariant Lagrangian. Their coupling to the $\rhoT$ gauge fields is introduced through the covariant derivatives
\begin{equation}
  D^\mu\xi_{L,R} \;=\; \partial^\mu\xi_{L,R} - i\gr\,\rhoT^\mu\,\xi_{L,R},
  \label{eq:covder}
\end{equation}
from which we define the vector and axial one-forms
\begin{equation}
  V^\mu = \frac{1}{2i}\bigl[(D^\mu\xi_L)\,\xi_L^\dagger
           + (D^\mu\xi_R)\,\xi_R^\dagger\bigr],
  \qquad
  A^\mu = \frac{1}{2i}\bigl[(D^\mu\xi_L)\,\xi_L^\dagger
           - (D^\mu\xi_R)\,\xi_R^\dagger\bigr].
  \label{eq:oneforms}
\end{equation}
These are the covariantized one-forms of the chiral fields with respect to the hidden local symmetry~\cite{Bando:1987br,Harada:2003jx}. 
The vector one-form $V^\mu$ is valued along the
unbroken ($T_L = T_R$) directions and contains the $\rhoT$ fields, while the axial
one-form $A^\mu$ is valued along the broken ($T_L = -T_R$) directions and contains the
techni-pions; both are $4\times4$ matrices expanded on the same
$\widetilde{\Lambda}^a$, all traces below being taken over the $4\times4$ flavour
space with $\tr[\widetilde{\Lambda}^a\widetilde{\Lambda}^b]=\tfrac{1}{2}\delta^{ab}$.
Both one-forms
transform homogeneously under the local symmetry, $V^\mu \to h\,V^\mu
h^\dagger$ and $A^\mu \to h\,A^\mu h^\dagger$, so that the traces of their
squares are invariant (see Appendix~\ref{app:HLS}). Using
these invariant combinations, the leading-order HLS Lagrangian
is~\cite{Bando:1985rf,Bando:1987br}
\begin{equation}
  \mathcal{L}_\mathrm{HLS}
  \;=\;
  \FPi^2\,\tr\bigl[A^\mu A_\mu\bigr]
  \;+\;
  a\,\FPi^2\,\tr\bigl[V^\mu V_\mu\bigr]
  \;-\;
  \frac{1}{2}\tr\bigl[\rhoT^{\mu\nu}{\rhoT}_{\mu\nu}\bigr],
  \label{eq:HLS}
\end{equation}
with field strength
$\rhoT^{\mu\nu} = \partial^\mu\rhoT^\nu - \partial^\nu\rhoT^\mu
- i\gr[\rhoT^\mu,\rhoT^\nu]$. The first term gives the canonically normalized
techni-pion kinetic term. The second term encodes the $\rhoT$ mass,
$\MrhoT^2 = a\,\gr^2\FPi^2$, together with its interactions with the
techni-pions. The third term gives the $\rhoT$ kinetic term and
self-interactions.

The decay phenomenology is governed by the vector one-form. Through the
structure constants $f^{abc}$ it mixes the $\rhoT$ fields with techni-pion
bilinears and thereby determines which decay channels are open for each
resonance. Expanding it to quadratic order in the fields makes this explicit
(Appendix~\ref{app:HLS}),
\begin{equation}
  V^\mu
  \;=\;
  -\gr\,\rhoT^{a\mu}\,\widetilde{\Lambda}^a
  \;+\;
  \frac{f^{bca}}{2\FPi^2}\,
  \partial^\mu\PiT^b\,\PiT^c\,\widetilde{\Lambda}^a
  \;+\;\mathcal{O}(\Pi^3),
  \label{eq:Vmu_expand}
\end{equation}
where the first term contains the propagating $\rhoT$ degrees of freedom and
the second term couples each $\rhoT^a$ to a pair of techni-pions
$(\PiT^b, \PiT^c)$ with a strength set by the structure constant $f^{bca}$.
Both terms are expanded along the unbroken (vector) directions, which in the $4\times4$ notation carry the same $\widetilde{\Lambda}^a$ as the broken ones. Inserting Eq.~\eqref{eq:Vmu_expand} into the $a\FPi^2\tr[V^\mu V_\mu]$ term of Eq.~\eqref{eq:HLS}, the cross term isolates the three-point interaction
\begin{equation}
  \mathcal{L}_{\rhoT\PiT\PiT}
  \;=\;
  g_{\rhoT\PiT\PiT}\,f^{abc}\,\rhoT^{a\mu}\,\PiT^b\,\partial_\mu\PiT^c,
  \qquad
  g_{\rhoT\PiT\PiT} = \frac{a\,\gr}{2},
  \label{eq:rhopipi}
\end{equation}
with the structure constants $f^{abc}$ selecting which pion pairs each
$\rhoT^a$ couples to. The derivation is given in Appendix~\ref{app:HLS}. From this vertex the partial decay width of $\rhoT^a$ into any pair $(\PiT^b, \PiT^c)$ follows as
\begin{equation}
  \Gamma\!\bigl(\rhoT^a \to \PiT^b\PiT^c\bigr)
  \;\propto\;
  g_{\rhoT\PiT\PiT}^2\,\bigl|f^{abc}\bigr|^2\,\frac{p_{bc}^3}{\MrhoT^2},
  \label{eq:width}
\end{equation}
where $p_{bc}$ is the momentum of either final-state pion in the $\rhoT$ rest frame, defined in Appendix~\ref{app:HLS}. The branching pattern is therefore determined by the group factors $|f^{abc}|^2$ together with the phase-space factors $p_{bc}^3$, with the coupling $g_{\rhoT\PiT\PiT}$ and all remaining factors being universal and cancelling in the ratio that defines the branching fraction. This vertex is thus the only dynamical input needed for the branching-ratio calculation of the next section.
\section{Decay Properties of the Spin-1 Resonances}
\label{sec:branching}

The dominant bosonic decay channels of the spin-1 resonances  will be the subject of consideration, with the  branching fractions only computed amongst them.
The partial width for the decay $\rhoT^a \to \PiT^b\PiT^c$ follows directly
from the vertex of Eq.~\eqref{eq:rhopipi} and is proportional to
$g^{2}_{\rhoT\PiT\PiT}\,|f^{abc}|^2$ times the $P$-wave phase-space factor,
Eq.~\eqref{eq:width}. The coupling $g_{\rhoT\PiT\PiT}$ and the mass
$\MrhoT$ are universal across all decay channels and cancel identically in the
ratio of partial widths that defines a branching fraction. What remains is
determined by the $\SU{4}$ structure constants $|f^{abc}|^2$ and the available
phase space of each channel, which is what we now compute.

To organize the calculation we classify the fifteen techni-pions into two
categories. The three light pions, denoted $\pi_{\rm HC}^V$, are the isotriplet states identified in the broken phase with the longitudinal $W^\pm$ and $Z^0$. The remaining twelve heavy pions, denoted $\PiT^H$, sit near the EW scale and are not absorbed as gauge boson modes.
The group-theoretic weight of each category is encoded in the group weights
$C_L^a$ and $C_H^a$, obtained by summing $|f^{abc}|^2$ over pairs of light
($\pi_{\rm HC}^V \sim W^\pm, Z^0$) and heavy ($\PiT^H$) pion indices, respectively,
together with the mixed weight $C_\mathrm{mix}^a$ for pairs of one light and
one heavy pion. The $\SU{4}$ Casimir relation fixes their sum for every
generator index $a$, $C_L^a + C_H^a + C_\mathrm{mix}^a = 4$, since the sum of
$|f^{abc}|^2$ over all pion pairs equals the quadratic Casimir of the adjoint
representation, $C_2(\mathrm{adj}) = 4$ for $\SU{4}$. Since all coupling and
mass factors cancel in the ratio, the branching fraction is a pure
group-theory and phase-space prediction with no dependence on the underlying
dynamics beyond the structure constants. The definitions of the group weights
and the derivation of the branching-ratio formula are given in
Appendix~\ref{app:phasespace}.

The decay pattern follows from the commutation relations between pairs of broken generators, summarized in Table~\ref{tab:broken-generators} in
Appendix~\ref{app:generators}.
Among these relations, four sectors of broken-generator pairs close onto the vector isotriplet $T^{V,(a)}$,
\begin{equation}
\begin{aligned}
  [K^{V,(a)},K^{V,(b)}] &= \frac{i}{\sqrt{2}}\,\eps^{abc}\,T^{V,(c)},
  &\qquad
  [K^{A,(a)},K^{A,(b)}] &= \frac{i}{\sqrt{2}}\,\eps^{abc}\,T^{V,(c)},
  \\[4pt]
  [K^{I,(i)},K^{I,(j)}] &= \frac{i}{\sqrt{2}}\,\eps_{ijk}\,T^{V,(k)},
  &\qquad
  [K^{II,(i)},K^{II,(j)}] &= \frac{i}{\sqrt{2}}\,\eps_{ijk}\,T^{V,(k)},
\end{aligned}
\label{eq:closure_main}
\end{equation}
where the generators $K^{V,(a)}$ are associated with the light pions
$\pi_{\rm HC}^V$, while $K^{A,(a)}$, $K^{I,(i)}$, and $K^{II,(i)}$ are associated with the heavy pions $\PiT^H$. No other commutator of two broken generators produces the isotriplet, so $T^{V,(a)}$ is the only unbroken generator that receives a light-light contribution, and the associated states $\rhoT^{\pm,0}$ are the only ones with di-EW-boson decay channels. The first relation of Eq.~\eqref{eq:closure_main} gives $C_L = 1$, the mixed weight vanishes for these states, and the sum rule then fixes $C_H = 3$ (Appendix~\ref{app:phasespace}), the three remaining relations of Eq.~\eqref{eq:closure_main} each contributing unity.
Accordingly, the isotriplet $\rhoT^{\pm,0}$, which transforms as the
$\mathbf{3}$ of $\SU{2}_V$ with $C_L = 1$ and $C_H + C_\mathrm{mix} = 3$, decays dominantly into $\pi^{V,(a)}_{\rm HC}\pi^{V,(b)}_{\rm HC}$ ($W^\pm, Z^0$), $\PiT^{A,(a)}\PiT^{A,(b)}$, $\PiT^{K^{\mathrm{I},(i)}}\PiT^{K^{\mathrm{I},(j)}}$ and $\PiT^{K^{\mathrm{II},(i)}}\PiT^{K^{\mathrm{II},(j)}}$, one channel for each of the four
relations of Eq.~\eqref{eq:closure_main}; the weights and dominant channels of all fifteen spin-1 states are summarized in Table~\ref{tab:decay_app} of Appendix~\ref{app:phasespace}.
Therefore, the branching ratio depends on the group weights together with the phase space,
\begin{equation}
  \mathrm{BR}\!\bigl(\rhoT^{\pm,0} \to W^\pm Z,\,W^+W^-\bigr)
  \;=\;
  \frac{\beta_{\pi_{\rm HC}^V}^3}{\beta_{\pi_{\rm HC}^V}^3 + 3\,\beta_{\PiT^H}^3},
  \label{eq:BR_final}
\end{equation}
where
\begin{equation}
  \beta_{\pi_{\rm HC}^V} = \sqrt{1 - \frac{4M_W^2}{\MrhoT^2}},
  \qquad
  \beta_{\PiT^H} = \sqrt{1 - \frac{4M_{\PiT^H}^2}{\MrhoT^2}},
\end{equation}
are the velocities of
the final-state pions in the light and heavy channels, and their cubes
$\beta^3_{\pi_{\rm HC}^V}$ and $\beta^3_{\PiT^H}$ are the corresponding $P$-wave
phase-space factors. For $\rhoT^\pm \to W^\pm Z$ the small $W$--$Z$ mass
splitting is negligible against $\MrhoT$ and a common $\beta_{\pi_{\rm HC}^V}$
with $M_W \simeq M_Z$ is used (Appendix~\ref{app:HLS}).
\begin{figure}[tbp!]
  \centering
  \begin{tabular}{@{}cc@{}}
    \includegraphics[width=0.48\linewidth]{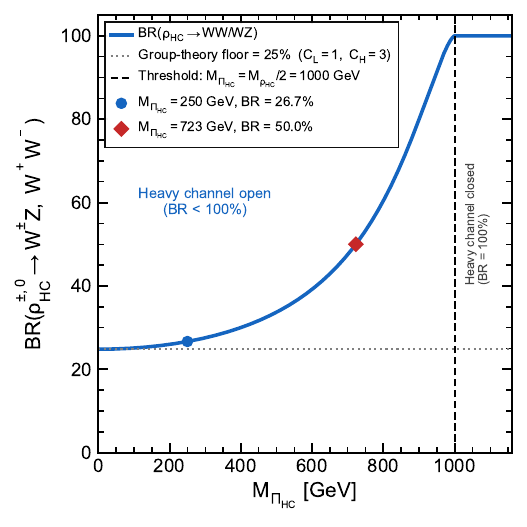} &
    \includegraphics[width=0.48\linewidth]{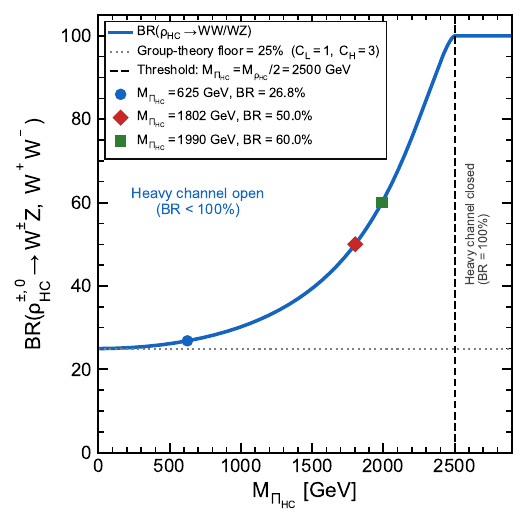} \\
  \end{tabular}
  \caption{Branching ratio
  $\mathrm{BR}(\rhoT^{\pm,0} \to W^{+}W^{-}/W^{\pm}Z)$ in $\%$ as a
  function of the heavy pion mass $M_{\PiT^H}$, for
  $\MrhoT = 2$~TeV (left) and $5$~TeV (right).}
  \label{fig:BR_plot}
\end{figure}
The branching ratio of Eq.~\eqref{eq:BR_final} has two clean limits. When the heavy pions are heavier than half the $\rhoT$ mass the heavy channel is kinematically closed, $\beta_{\PiT^H}=0$, and the branching ratio is unity. When $M_{\PiT^H} \ll \MrhoT$ both phase-space factors approach unity and the branching ratio tends to the group-theory value
$C_L/(C_L+C_H) = \tfrac{1}{4}$.
Figure~\ref{fig:BR_plot} shows the branching ratio between these limits, as a function of $M_{\PiT^H}$ for $\MrhoT = 2$~TeV (left) and $5$~TeV (right). As $M_{\PiT^H}$ approaches the threshold $\MrhoT/2$ from below, the heavy channel closes as $\beta_{\PiT^H}\to0$ and the branching ratio into $W^\pm Z$ and $W^+W^-$ rises continuously toward unity. For our benchmark spectra both decay channels are well above threshold. In the TC limit ($\LTC = 2\,\mathrm{TeV}$), Eqs.~\eqref{eq:mass_pi} and \eqref{eq:mass_rho} give $\MrhoT\approx2\,\mathrm{TeV}$ and $M_{\PiT^H}\approx250\,\mathrm{GeV}$. Away from the TC limit, larger $\FPi$ ($\LTC = 5\,\mathrm{TeV}$) gives $\MrhoT\approx5\,\mathrm{TeV}$ and $M_{\PiT^H}\approx625\,\mathrm{GeV}$. Both spectra give $\mathrm{BR}(\rhoT^{\pm,0}\to W^\pm Z,\,W^+W^-)\approx27\%$, only slightly above $\tfrac{1}{4}$, since the ratio $M_{\PiT^H}/\MrhoT$ is fixed by the common scaling with $\FPi$. The phase-space difference between the EW bosons and the heavy pion is a small correction to the group weights.

Since only the isotriplet receives a light-light contribution, the remaining
twelve spin-1 states have $C_L^a = 0$ and do not decay into di-EW-boson final
states. Their decays proceed instead through the mixed channels
$\pi_{\rm HC}^V\PiT^H$ and the heavy channels $\PiT^H\PiT^H$, with weights
satisfying $C_H^a + C_\mathrm{mix}^a = 4$. The two classes lead to different
final states, $W/Z$ plus a heavy pion versus a pair of heavy pions, with
different kinematic thresholds, $M_{W/Z} + M_{\PiT^H}$ and $2M_{\PiT^H}$,
respectively, and the dominant channels of each state are summarized in
Table~\ref{tab:decay_app}, with the complete set of commutation relations for
all fifteen unbroken generators grouped in Table~\ref{tab:commutators_app} of
Appendix~\ref{app:generators}.

One of the heavy pions (isosinglet state \cite{Ma:2015gra,Wu:2017iji})  decays through the Wess--Zumino--Witten (WZW)
anomaly~\cite{Wess:1971yu,Witten:1983tw} into pairs of EW gauge bosons,
$\PiT^H \to VV'$ with $V,V' = W^\pm, Z, \gamma$. The other pNGBs are protected by a $\mathbb{Z}_2$ parity, which can only be broken by the partial compositeness couplings. Therefore, partial
compositeness operators that generate the SM fermion
masses~\cite{Kaplan:1991dc} also induce couplings of the pNGBs to fermion
pairs proportional to the fermion mass, so that the neutral pions decay into $t\bar{t}$ while the charged pions decay into  $t\bar{b}$~\cite{Wu:2017iji,Belyaev:2016ftv}.  The mixed channels
$\pi_{\rm HC}^V\PiT^H$ thus lead to three-boson final states, or to $W/Z$ plus a
fermion pair, while the heavy channels $\PiT^H\PiT^H$ lead to four-boson
final states, or to bosons along with fermion pairs. These multi-boson
cascades arise both from the twelve states with vanishing light-light weight,
$C_L^a = 0$, whose decays proceed exclusively through such channels, and from
the heavy channel of the isotriplet $\rhoT^{\pm,0}$ itself. In this work we
focus on the clean di-boson decays of the isotriplet, leaving the richer
multi-boson and mixed boson--fermion final states for future study.

It must be stressed that the estimation of the branching fraction among the
bosonic decay modes constitutes a fair assumption in the context of this study. As the $\rhoT$ is well above all SM fermion pair thresholds, the
channel $\rhoT \to f\bar{f}$ is kinematically open.
However, SM fermions carry no hypercolor charge and therefore do not
couple directly to the $\rhoT$ through its strong hypercolor current.
The $\rhoT$, therefore, couples to them only
via vector meson dominance (VMD), through mixing with the $W$ and $Z$ bosons
with a small mixing angle
$\epsilon \sim g/g_{\rhoT}$, and it is through this mixing that
the $\rhoT$ inherits a coupling to the SM fermion vector
current~\cite{Casalbuoni:1985kq,Lane:1999uh,Lane:2002sm,PhysRevD.92.085019,Pappadopulo_2014,Contino:2011np}.
This is the exact analogue of how the QCD $\rho^0$ acquires its $e^+e^-$
partial width through $\rho^0$--$\gamma$ mixing in
VMD~\cite{Sakurai:1960ju,Gell-Mann:1961jim,Kroll:1967it,Bando:1985rf,Bando:1987br}, but here the suppression factor is
$(g/g_{\rhoT})^2 \ll 1$ rather than being controlled by the electromagnetic coupling $\alpha$.
In both cases the fermion decay rate is parametrically smaller than the
di-boson rate, which proceeds at full HLS coupling strength
$g_{\rhoT\PiT\PiT}$ without any such
suppression~\cite{Bando:1987br,Harada:2003jx,Lane:2009ct,Pappadopulo_2014}.
The di-boson channel therefore dominates the $\rhoT^{\pm,0}$ decay width,
which is the basis of the collider phenomenology pursued in this paper.

The $\omT$, associated with the singlet $\UN{1}_V$ current, is a
qualitatively distinct case and does not decay through the HLS vertex at all.
In QCD the isosinglet $\omega$ meson decays dominantly into three pions, and
this process is not governed by the HLS Lagrangian but rather by the
WZW anomaly term~\cite{Wess:1971yu,Witten:1983tw}, which encodes the chiral anomaly of
the underlying gauge theory in the effective meson Lagrangian.
The decay $\omega \to \pi^+\pi^-\pi^0$ proceeds through the anomalous
$\omega\rho\pi$ coupling that arises from this topological term and is
forbidden by the normal parity selection rules of the HLS
sector~\cite{Bando:1987br,Harada:2003jx,PhysRevD.30.594}.
By the same analogy, the $\omT$ in our hypercolor sector decays into three
EW bosons $W^+Z^0W^-$ through the WZW anomaly, in direct
correspondence with the QCD $\omega \to \pi^+\pi^-\pi^0$ decay, as discussed
in~\cite{Chivukula:1989rn}. This channel is governed entirely by the
topological structure of the hypercolor theory.
More precisely, the topological term is not restricted to the $W^+Z^0W^-$ channel, since it couples the $\omT$ to every pion triple $(\PiT^a,\PiT^b,\PiT^c)$
with a non-vanishing antisymmetric structure constant $f^{abc}$, so that
mixed triples $\pi_{\rm HC}^V\pi_{\rm HC}^V\PiT^H$ and $\pi_{\rm HC}^V\PiT^H\PiT^H$, as
well as all-heavy triples $\PiT^H\PiT^H\PiT^H$, also contribute. The corresponding weights follow from the commutation relations of Table~\ref{tab:broken-generators}, where each non-vanishing entry corresponds to one three-body channel, as described in Appendix~\ref{app:HLS}. Among these channels, the final state $W^+Z^0W^-$ is the dominant one, since the anomalous vertex carries three derivatives and the resulting matrix element grows with the momenta of the final-state particles, so the width is controlled by the available three-body phase space. The phase space is largest for the lightest final state, and it decreases progressively as one, two, and then three EW bosons are replaced by heavy techni-pions. 
The same phase-space hierarchy makes $\omega \to \pi^+\pi^-\pi^0$ the dominant decay of the isosinglet vector in QCD. A quantitative determination of the subleading weights requires a dedicated three-body analysis, which we leave for future work, while the
$W^+Z^0W^-$ channel considered here is both the dominant one and the direct counterpart of the QCD decay that motivates the analogy.

The $\etaT$, the pseudoscalar singlet of the broken $\UN{1}_A$, follows a
different decay pattern. Because it is a bound state of the anomalously broken axial
symmetry, its coupling to the light SM fermions is proportional
to their mass divided by $\FPi$, and its dominant decay channel in the TC limit is into a top-antitop pair $\etaT \to t\bar{t}$~\cite{DIVECCHIA1980247,PhysRevD.96.075040,Holdom:1995hd,Hill:2002ap}.
This decay proceeds through interactions that
also generate the top quark mass, and the large top Yukawa coupling makes it
the preferred channel over lighter fermion pairs or boson
pairs~\cite{PhysRevD.96.075040,Chivukula:2011ue}.
The $\etaT \to t\bar{t}$ final state therefore provides a complementary
search channel at the FCC-hh to the di-boson channel of $\rhoT$, and its
phenomenology will be discussed in Section~\ref{sec:pheno}.
The $\etaT$ also couples to
pairs of EW gauge bosons through the topological anomaly, giving the
channels $\etaT \to \gamma\gamma,\, Z\gamma,\, ZZ,\, W^+W^-$ with
coefficients fixed by the anomaly of the $\UN{1}_A$ current, as studied for
states of this type in Refs.~\cite{Belyaev:2016ftv,PhysRevD.96.075040,Hill:2002ap,Cacciapaglia:2019bqz}. For $M_{\etaT}$ above the
$t\bar t$ threshold the mass-proportional coupling keeps $t\bar t$ the
dominant mode, while the anomalous di-boson channels, in particular $\gamma\gamma$ and $Z\gamma$, remain clean subleading signatures at a hadron
collider.

\section{Model Spectroscopy at Future Colliders}
\label{sec:pheno}
The main focus of this section is the production of composite EW bosons and
heavy resonances through the fragmentation and hadronization of hyperquarks, and
their reach at a future $100\,\mathrm{TeV}$ collider. However, these spin-1 resonances
are already constrained by existing data. We therefore begin by asking
what limits the LHC currently places on them and how far its reach extends,
before turning to the hadronization-based strategy at the FCC-hh~\cite{FCC:2018byv} that is developed in the subsections that follow.

\FloatBarrier
\subsection{Imprints at the LHC}
\label{sec:LHC}
The  production  of these resonances through hyperquark hadronization in the Drell-Yan process ($pp\rightarrow q_{\rm HC}\bar q_{\rm HC}$) presents a universal process by means of which every resonance in the spectrum can be accessed. This  window is highly unlikely at the LHC owing to the limited partonic centre-of-mass energy. However, LHC searches for these states, and the $\rho$ in particular, are possible by means of either direct or associated production. It is therefore instructive to discuss their reach by classifying them on the basis of their production as follows:
\begin{itemize}
    \item  Drell-Yan (DY)/VBF: A di-boson final state for a heavy vector resonance  is  favoured in many BSM scenarios. The coupling ($g_{\rho_{\rm HC}}$) of the $\rho_{\rm HC}$ to the SM bosons can be expressed in terms of the corresponding QCD value ($g_{\rho\pi\pi}$) as \cite{Piai:2010ma,Pappadopulo_2014,Baker_2022,PhysRevD.92.085019}
    \begin{equation}
        g_{\rho_{\rm HC}}= c_Hg_{\rho\pi\pi}\sqrt{\frac{3}{N_{\rm HC}}},
    \end{equation}
   where $g_{\rho\pi\pi}\sim 5.5$ and $c_H$ is a model-dependent $\mathcal{O}(1)$ parameter.
   The coupling of the vectors  to the fermions is due to the mixing with the gauge bosons and can be expressed as
   \begin{equation}
       g^{\rm DY}_{F} \;=\; c_{F}\,\frac{g^{2}}{g_{\rhoT}}\,\sqrt{\frac{\NHC}{3}},
   \end{equation}
   where $c_F$ is also a model-dependent $\mathcal{O}(1)$ parameter. Using this, the production cross-section due to both the DY and the VBF mode is given as 
    \begin{equation}
     \sigma^{\rm DY}(\NHC) \;=\; \frac{\NHC}{3}\;\sigma^{\rm DY},
     \qquad
     \sigma^{\rm VBF}(\NHC) \;=\; \frac{3}{\NHC}\;\sigma^{\rm VBF}.
  \label{eq:xsec-scaling}
  \end{equation}
  where $\sigma^{\rm DY} \propto c^{2}_{F}\,g^{4}/g^{2}_{\rhoT}$ and $\sigma^{\rm VBF} \propto (c_{H} g_{\rho\pi\pi})^{2}$. The masses of the resonances follow $m_{\rho_{\rm HC}}\sim\Lambda_{\rm HC}\sim4\pi v/(\sqrt{N_{\rm HC}}\sin\theta)$, where $\theta$ is the vacuum alignment angle. Thus the DY and VBF have a contrasting dependence on $N_{\rm HC}$. It is instructive to see the constraints on the model  in the  QCD-like limit ($N_{\rm HC}=3$ and $\theta=\pi/2$) with $c_{H,F}=1$.
    The yellow band of Figure~\ref{fig:LHCDYVFB} gives the projected $90\%$ upper bound expected for the DY+VBF combined at the HL-LHC~\cite{ATLAS:2020fry,CMS:2026fjw,ATLAS:2024qvg,ATLAS:2018ocj}. The red (blue) solid line gives the computed cross-sections for the DY (VBF), for the QCD-like model. 
    The left (right) panel corresponds to the charged (neutral)  heavy vector.  Thus the expected upper bounds on the charged and neutral $\rhoT$ masses, through the DY (VBF), are $\sim 4\, \textrm{TeV}$ ($\sim 2.2$ TeV) and $\sim 3.6\, \textrm{TeV}$ ($\sim 2.0$~TeV), respectively, as shown in Figure~\ref{fig:LHCDYVFB}.
  \begin{figure}[htb!]
    \begin{tabular}{cc}
     \includegraphics[width=0.5\linewidth]{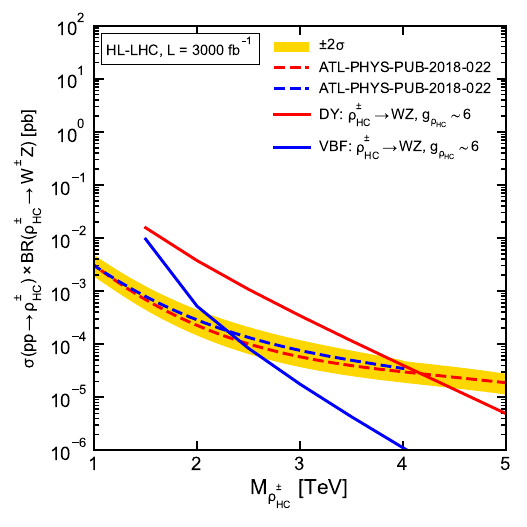}    &  \includegraphics[width=0.5\linewidth]{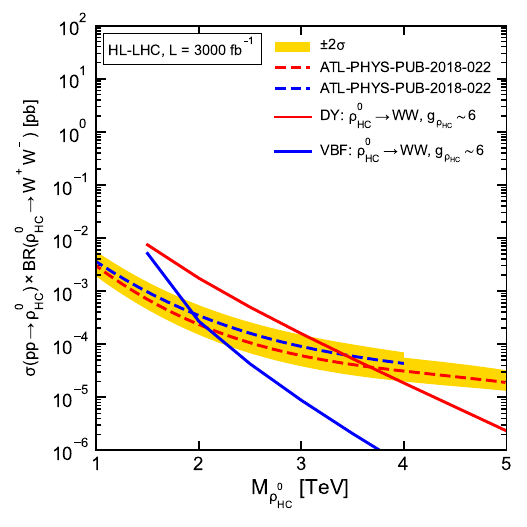}\\       
    \end{tabular}
        \caption{
Leading-order cross-section times branching ratio for the charged
$\rhoT^{\pm} \to W^{\pm}Z$ (left) and neutral $\rhoT^{0} \to W^{+}W^{-}$
(right), for DY (solid red line) and VBF (solid blue line) production. The
HL-LHC projections~\cite{ATLAS:2018ocj} are shown as dotted red (DY) and
dotted blue (VBF) lines, with the yellow band giving the expected $90\%$
upper bound for the combined DY+VBF search.}
        \label{fig:LHCDYVFB}
    \end{figure}   
  \item Associated production: A pair of these resonances can be produced by means of a process like $pp\rightarrow \rhoT~\PiT$. These bosons could then decay fermionically or through one of the modes in Table~\ref{tab:decay_app}. The bosonic mode is challenging at the LHC as the lack of boost of the EW bosons could lead to a high multiplicity of jets, making it difficult to differentiate from the multi-jet QCD background. The analysis at FCC-hh would be interesting as the boost would render the identification of the final-state bosons as fat-jets, eventually leading to the reconstruction of the resonances. 
\end{itemize}

The strongest bound is expected from DY production, which is proportional to $c_F^2$; hence the reach at the HL-LHC could be reduced if $c_F$ is significantly smaller than unity. Hence, it is premature to conclude that the TC limit, where we expect a $\rhoT$ mass around $2$~TeV, is excluded.

\subsection{Production from hadronization at FCC-hh}
\label{sec:pheno:setup}

In the previous Section~\ref{sec:LHC} we discussed the single production of the spin-1 heavy
resonances through the DY, VBF, and associated production channels at the LHC, together with the resulting upper bounds. These channels remain sensitive at a
future $100\,\mathrm{TeV}$ collider, where they probe masses up to
$M_{\rhoT} \sim 30\,\mathrm{TeV}$ for the QCD-like model benchmark point~\cite{Pappadopulo_2014}. In light of the HL-LHC constraints discussed for the QCD-like model, we consider the following two benchmarks:
\begin{itemize}
    \item \textbf{Benchmark 1} A QCD-like model with $c_H=c_F=1$ places a strong upper bound on the masses of the resonances. This can be alleviated by choosing $c_H,c_F<1$, such that the bounds are consistent with $\Lambda_{\rm HC}\sim 2$ TeV, which represents the exact QCD-like scaling ($\theta=\pi/2$), with $m_{\rhoT,\omT,\etaT}\sim 2$ TeV.
    Possible departures from $c_H,c_F=1$ have been discussed in specific models in~\cite{PhysRevD.92.085019}.
  \item \textbf{Benchmark 2} Departing from the pure TC vacuum ($\theta=\pi/2$) correspondingly increases the confinement scale as $\LambdaT\rightarrow\LambdaT/\sin\theta$. In this case we consider $\LambdaT=5$ TeV, with the resonances at the same masses. 
\end{itemize}
A crucial fact, as noted in Eq.~\eqref{eq:rho_iso_currents}, is that the currents for the resonance of interest are independent of $\theta$. This  renders a universality for the search strategy across both benchmarks.
As a complementary strategy to direct production, we propose a model-independent production channel for techni-hadrons, based on the fragmentation and hadronization of hyperquarks ($\q$) at the FCC-hh, for Benchmark~1\,(2).

The hyperquark pairs are produced in
the DY process $pp \to \q\bar{q}_{\rm HC}$, and these final-state hyperquarks fragment and hadronize into a spray of techni-hadrons, in precise analogy with the way a pair of SM quarks $q\bar{q}$ in QCD fragments and hadronizes into ordinary hadrons, such as pions and $\rho$ mesons, through the
Lund string mechanism~\cite{Andersson:1983ia,Andersson:1983jt}. This analogy was used to propose the composite EW boson signature at the FCC-hh and at muon
colliders in our earlier work~\cite{Cacciapaglia:2025multiboson}. We generate $200{,}000$ hard-process signal events at leading order with
\textsc{MadGraph5\_aMC@NLO}~\cite{Alwall:2014hca}\footnote{We use the NNPDF2.3LO parton
distribution set~\cite{Ball:2012cx}, and the factorization and renormalization scales are set dynamically in
\textsc{MadGraph5\_aMC@NLO}~\cite{Alwall:2014hca}, evaluated event by event from the kinematics of the hard process. For the production of heavy hyperquarks these scales lie close to the constituent hyperquark mass, $\mathcal{O}(2\,\mathrm{TeV})/\mathcal{O}(5\,\mathrm{TeV})$ for Benchmark~1/2.}, using the UFO model file~\cite{Degrande:2011ua} generated with
\textsc{FeynRules}~\cite{Alloul:2013bka}. The events are then passed to the Hidden Valley (HV) module~\cite{Carloni:2011kk} of
\textsc{Pythia}~8.312~\cite{Bierlich:2022pfr} for showering and hadronization. We set the fragmentation parameters of the HV module such that, with the appropriate normalization, they correspond to the SM QCD values. The full details of the HV module and its fragmentation settings are discussed in our earlier work~\cite{Cacciapaglia:2025multiboson}. For Benchmark 1\,(2), the minimum partonic centre-of-mass energy $\sqrt{\hat{s}}_\mathrm{min} = 10 \,(12)\,\mathrm{TeV}$ is imposed at the hard-scattering level in \textsc{MadGraph5\_aMC@NLO}, and the leading-order
cross-section is $\sigma_{\rm sig} = 667.9\,(182.7)\,\mathrm{ab}$ for the number of
hypercolors $\NHC = 3$ and the number of flavours $N_f = 4$. The hadronization
of the hyperquark pairs gives a rich spectrum of spin-0 and spin-1
techni-hadrons, the analogues of the QCD pions, $\rho$, $\omega$ and $\eta$
mesons. The techni-pion triplet $\pi^{\pm,0}_{\rm HC}$ is identified with the longitudinal components of the EW bosons $W^\pm$ and $Z^0$, and the fraction of events containing at least one EW boson from string fragmentation is $27\,(18)\%$. Additional EW bosons are produced in the decays
of the heavy spin-1 resonances. Our signals are these EW bosons together with
the QCD-like heavy resonances, the isotriplet vectors $\rhoT^{\pm,0}$, the
isosinglet vector $\omT$ and the pseudoscalar
$\etaT$, and the corresponding event fractions are $14.6\,(9.4)\%$, $7.1\,(3.0)\%$ and $15.1\,(5.2)\%$. These fractions are small because the fragmentation probability
is shared among the sixteen spin-1 states and the techni-pions. The heavy
techni-pions $\Pi^{H}_{\rm HC}$ and the remaining spin-1 states are not
considered further. Detector response is modelled with the FCC-hh card of \textsc{Delphes}~\cite{deFavereau:2013fsa}. 

\FloatBarrier
\subsection{Object reconstruction and event topology}
\label{sec:pheno:collider}
\begin{figure}[htbp]
    \centering
    \begin{tabular}{@{}cc@{}}
        \includegraphics[width=0.48\textwidth]{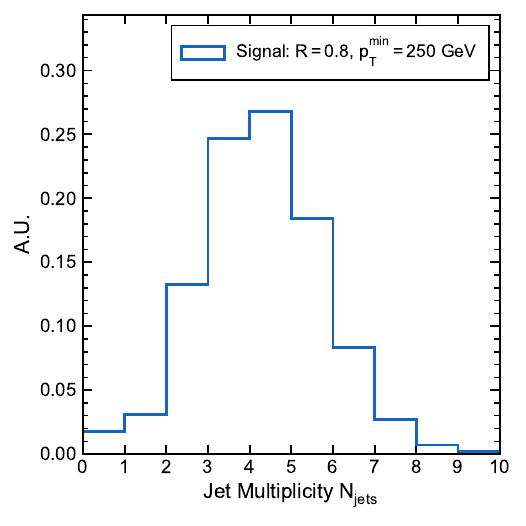} &
        \includegraphics[width=0.48\textwidth]{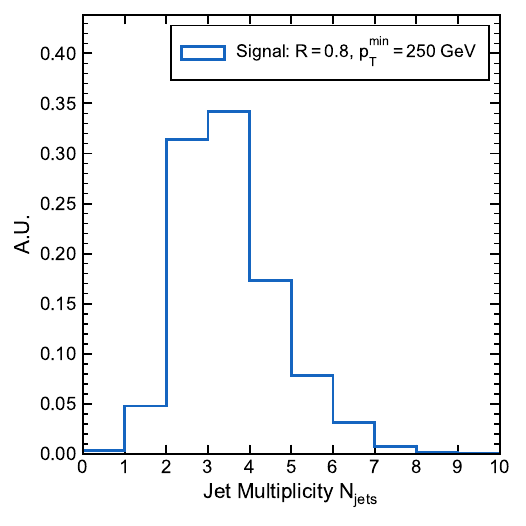} \\
    \end{tabular}
    \caption{Jet multiplicity distribution in the multi-EW-boson final state, for jets clustered with $R = 0.8$ and selected with
    $p_{T} > 250$~GeV and $|\eta| \leq 2.5$, for Benchmark~1 (left) and Benchmark~2 (right), generated at $\sqrt{\hat{s}}_{\min} = 10$~TeV and $12$~TeV respectively, for the minimal observable with $\Nf = 2$ and $\NHC = 3$.}
    \label{fig:mult}
\end{figure}

In this study we focus on the hadronic decays of the EW gauge bosons (and tops), which we identify as single ``fat'' jets. These
large-$R$ jets are reconstructed with \textsc{FastJet}~\cite{Cacciari:2011ma}
and clustered using the anti-$k_T$ algorithm~\cite{Cacciari:2008gp} with a
radius parameter $R=0.8$. The reconstructed jets are required to have $p_T > 250\,\mathrm{GeV}$ and $|\eta|\leq 2.5$. This radius is large enough
to capture the full hadronic decay products of a boosted EW boson or top quark
within a single fat jet, as expected for the bosons originating from a
$2\,(5)\,\mathrm{TeV}$ resonance and those produced directly from hyperquark fragmentation~\cite{Cacciapaglia:2025multiboson}. 
The number of
reconstructed jets per event peaks at four (three) for Benchmark 1 (2), as
shown in the left (right) panel of Figure~\ref{fig:mult}.
The jet multiplicity tracks the number of EW bosons and top quarks in the
event. The light techni-pions, identified with the EW bosons, are produced
either directly in the hyperquark fragmentation or in the decays of the spin-1
resonances, and each hadronically decaying boson
forms a single $W/Z$ jet after
clustering. The bosons produced directly in the fragmentation set the baseline
of this count. On top of this, the $\rhoT$ decays into a pair of EW bosons and
the $\omT$ into three, while the spin-0 $\etaT$ decays into a top-antitop
pair, with each top decaying predominantly into a $W$ boson and a $b$ quark;
for a hadronically decaying $W$ the boosted top is captured within a single
top jet. These di-boson, tri-boson and top-pair contributions from the
resonance decays push the jet count of signal events into the three-to-four
range, whereas SM multi-boson production, which lacks any resonant source of
extra bosons, populates lower jet multiplicities on average.

To optimize the identification of final-state jets, corresponding to $W/Z$ and top quarks, we employ grooming, as detailed in Appendix~\ref{app:kinematics}. 
The observed structures motivate the two mass windows used throughout the
analysis to define the physics objects: an EW boson tag with a mass range of
$60$--$110\,\mathrm{GeV}$, which selects hadronic $W$ and $Z$ decays
reconstructed as a single large-$R$ jet, and a top-jet mass window of
$150$--$350\,\mathrm{GeV}$, which captures the three-pronged topology of
$t\to Wb\to q\bar{q}'b$.

The irreducible SM background common to the spin-1 isotriplet ($\rhoT$) and
isosinglet ($\omT$) searches, and to the inclusive EW boson excess, is the
direct production of SM bosons, $pp \to XY,\,XYZ$ with
$X,Y,Z \in \{W^\pm, Z\}$. The higher multiplicities are populated through the
pair production of off-shell SM bosons, which subsequently decay into
multi-bosons, followed by their hadronic decays~\cite{Cacciapaglia:2025multiboson}. For the spin-0 $\etaT$ the
relevant background is instead associated top-pair production,
$pp \to t\bar{t}W^{\pm}/Z/H$. Both backgrounds are generated at leading order
with \textsc{MadGraph5\_aMC@NLO}~\cite{Alwall:2014hca} and showered with
\textsc{Pythia}~8.312~\cite{Bierlich:2022pfr}, with the jet reconstruction and
clustering performed as for the signal. At
$\sqrt{\hat{s}}_\mathrm{min} = 10\,(12)\,\mathrm{TeV}$ the cross-sections are
$\sigma_{\rm bkg} = 1786.7\,(844.0)\,\mathrm{ab}$ for the multi-boson
background and $11654.0\,(4524.7)\,\mathrm{ab}$ for the associated top-pair
background.

\begin{figure}[htbp]
    \centering
    \begin{tabular}{@{}cc@{}}
        \includegraphics[width=0.48\textwidth]{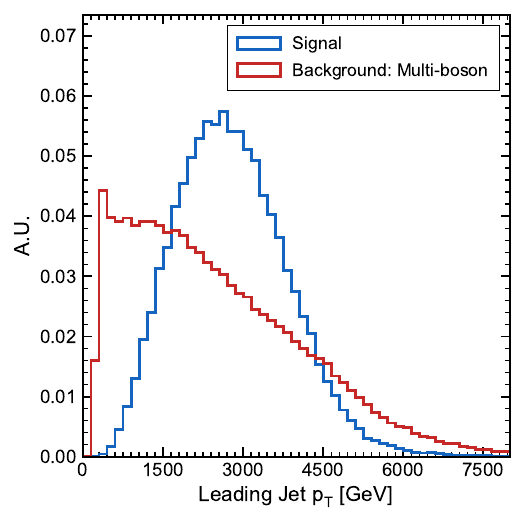} &
        \includegraphics[width=0.48\textwidth]{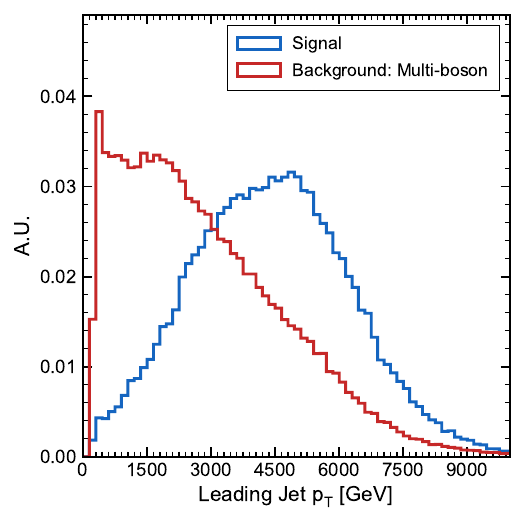} \\
    \end{tabular}
    \caption{Leading EW boson jet $p_{T}$ distribution for the signal (blue)
    and the SM multi-boson background (red), for Benchmark~1 (left) and Benchmark~2 (right), generated with $\sqrt{\hat{s}}_{\min} = 10$~TeV and $12$~TeV
    respectively.}
    \label{jetpT}
\end{figure}

The transverse momentum, $p_T$, provides an additional discriminator between
the signal EW bosons and the SM multi-boson background. The
leading-jet $p_T$ distributions are shown in Figure~\ref{jetpT} for
Benchmark~1 and Benchmark~2 in the left and right panels, respectively. The signal EW bosons, which originate from hyperquark fragmentation and hadronization and are also produced in the decays of the heavy resonances of mass $2\,(5)\,\mathrm{TeV}$, have $p_{T}$ in the TeV range. They therefore populate the hard and
central region of phase space. In contrast, the SM background bosons are
produced through the underlying hard-scattering processes and subsequent
EW radiation and decays, resulting in a softer $p_T$ spectrum and a
larger contribution at forward rapidities.

\begin{figure}[htbp]
    \centering
    \begin{tabular}{@{}cc@{}}
        \includegraphics[width=0.48\textwidth]{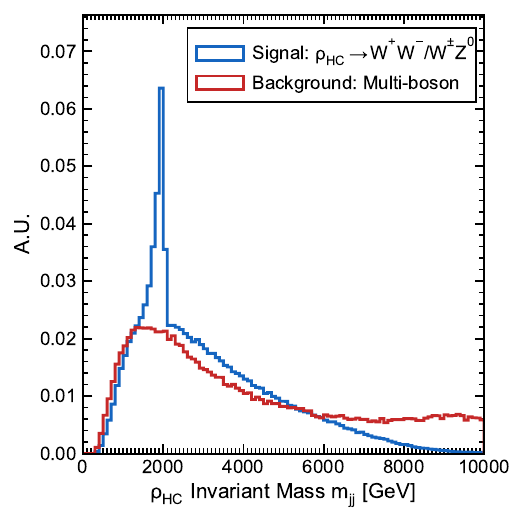} &
        \includegraphics[width=0.48\textwidth]{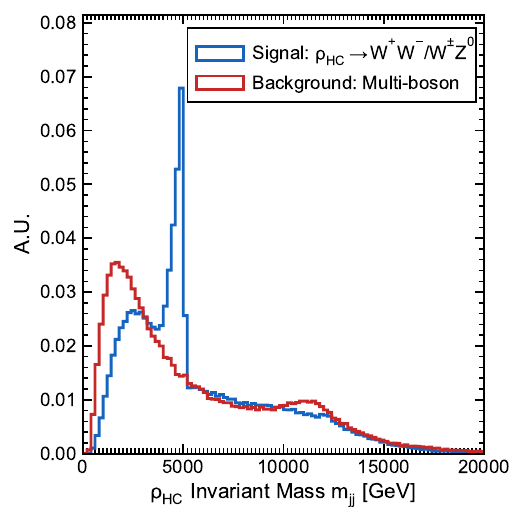} \\
    \end{tabular}
    \caption{Dijet invariant mass distribution for the isotriplet
    $\rhoT$ signal (blue) and the SM multi-boson background (red), at
    $M_{\rhoT} = 2$~TeV (left) and $5$~TeV (right), generated with $\sqrt{\hat{s}}_{\min} = 10$~TeV and $12$~TeV
    respectively.}
    \label{fig:8}
\end{figure}

\paragraph{Isotriplet $\rhoT$ search.}
The isotriplet $\rhoT$, with a mass of $2\,(5)\,\mathrm{TeV}$ for Benchmark~1~(2), is produced with $p_T$ comparable to its mass. In analogy with $\rho \to \pi\pi$ in QCD, it decays into a pair of longitudinal EW bosons, $\rhoT^{\pm,0} \to W^{+}W^{-}/W^{\pm}Z^{0}$, with a branching ratio of about $27\%$. Since the resonance is boosted, the two daughters share its momentum asymmetrically, see  Appendix~\ref{app:kinematics}. They  remain nevertheless well separated, with an angular distance $\Delta R \sim 2$, so that each $\rhoT$ gives two resolved jets rather than a single merged jet. The signal is therefore sought as a bump in the invariant mass of two boson-tagged jets rather than in the jet mass distribution. These EW bosons are themselves boosted, so their hadronic decays form collimated jets whose groomed mass lies close to the EW boson mass. The $\rhoT$ invariant mass is reconstructed as
\begin{equation}
  m_{\rhoT,(ik)}^{\mathrm{reco}} = \sqrt{\left(p_{j_i} + p_{j_k}\right)^{2}},
  \qquad 0 \leq i < k \leq n-1 ,
  \label{eq:mrho_reco}
\end{equation}
from events with a jet multiplicity $n \geq 3$, using boson-tagged jets with a groomed mass in the range $60$--$110\,\mathrm{GeV}$. In addition to the EW bosons from the $\rhoT$ decay, highly boosted EW bosons are produced directly in the string fragmentation, and these also carry $p_T$ in the TeV range. The leading jet can therefore originate from either source, so pairing only with respect to the leading jet would not give as pure a sample of $\rhoT$-like pairs in the resonance mass region. We therefore consider all $\binom{n}{2}$ dijet combinations, so that every potential $\rhoT$-like pair in an event is included. The distribution, shown in Figure~\ref{fig:8}, exhibits a $\rhoT$ resonance peak around its generated mass, together with a smoothly falling continuum from the SM EW boson background, which is reconstructed in the same way as the signal.

\begin{figure}[htbp]
    \centering
    \begin{tabular}{@{}cc@{}}
        \includegraphics[width=0.48\textwidth]{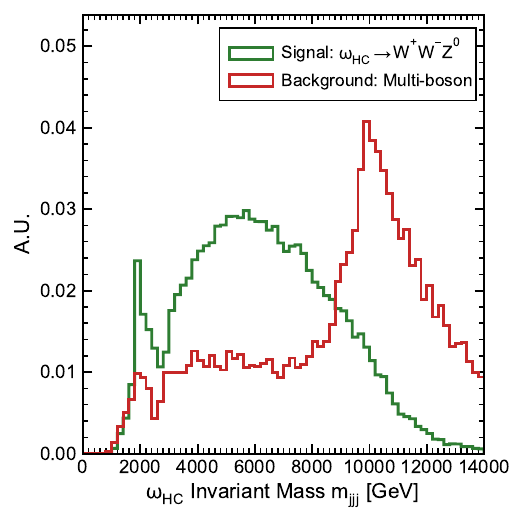} &
        \includegraphics[width=0.48\textwidth]{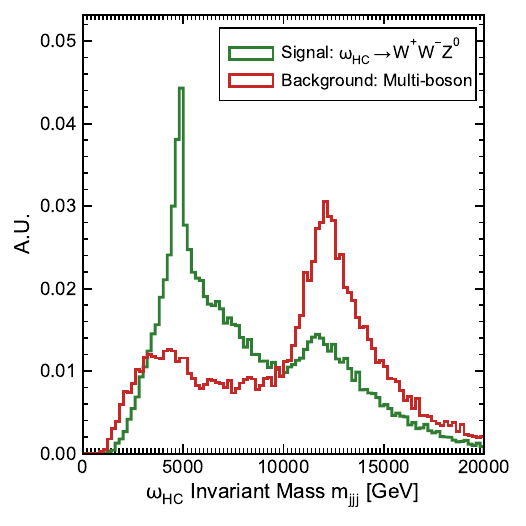} \\
    \end{tabular}
    \caption{Trijet invariant mass distribution for the isosinglet
    $\omT$ signal (green) and the SM multi-boson background (red), at
    $M_{\omT} = 2$~TeV (left) and $5$~TeV (right), generated with $\sqrt{\hat{s}}_{\min} = 10$~TeV and $12$~TeV respectively.}
    \label{fig:11}
\end{figure}

\paragraph{Isosinglet $\omT$ search.}
The isosinglet $\omT$, with a mass of $2\,(5)\,\mathrm{TeV}$ for Benchmark~1~(2), is likewise produced with $p_T$ comparable to its mass. Its leading decay is into three EW bosons, $\omT \to W^{+}Z^{0}W^{-}$, as in the QCD decay $\omega \to 3\pi$. As for the $\rhoT$, the resonance momentum is distributed asymmetrically among the three daughters, see  Appendix~\ref{app:kinematics}, and the three bosons remain separated, with $\Delta R \sim 2$, so each one gives its own resolved jet, and each is boosted enough that its hadronic decay products merge into a single large-$R$ jet. The $\omT$ invariant mass is therefore reconstructed from trijet combinations of boson-tagged jets,
\begin{equation}
  m_{\omT,(abc)}^{\mathrm{reco}} = \sqrt{\left(p_{j_a} + p_{j_b} + p_{j_c}\right)^{2}},
  \qquad 0 \leq a < b < c \leq n-1 ,
  \label{eq:momega_reco}
\end{equation}
from events with a jet multiplicity $n \geq 3$ and at least three boson-tagged jets with a groomed mass in the range $60$--$110\,\mathrm{GeV}$. All $\binom{n}{3}$ combinations are retained, except that a veto removes any trijet combination containing a pair whose invariant mass lies within $\pm 200\,\mathrm{GeV}$ of the $\rhoT$ mass, $m_{\rhoT} = 2\,(5)\,\mathrm{TeV}$, so that the two searches do not share the same jets. The distribution is shown in Figure~\ref{fig:11}.
The reconstructed spectrum shows a peak at the generated $\omT$ mass on top of a broad continuum, and the continuum has two distinct origins. The first is combinatorial: with $\binom{n}{3}$ triplets per event, most combinations group jets that do not all descend from the same resonance, and these populate the spectrum smoothly rather than at the resonance mass. The second is the generation cut: the requirement on $\sqrt{\hat{s}}_{\min}$ places a floor on the partonic energy of the event, so the reconstructed mass is driven well above the resonance mass and the continuum persists far beyond the peak region. Both effects are present in the signal and in the SM multi-boson background, which is reconstructed in the same way.

The peak itself is less pronounced at $2\,\mathrm{TeV}$ than at $5\,\mathrm{TeV}$, and this follows directly from the $p_T$ ordering of the daughters. At $2\,\mathrm{TeV}$ the soft boson carries $p_T$ of approximately $340\,\mathrm{GeV}$, with its lower tail extending below the jet selection criterion $p_T^{\min} = 250\,\mathrm{GeV}$ (see Figure~\ref{fig:10} in Appendix~\ref{app:kinematics}), so in those events the third jet is not reconstructed and the complete triplet is lost, leaving only the combinatorial background in the window. The threshold cannot be lowered to recover them, since it is fixed by the jet radius: below $2m_V/R \simeq 200$--$230\,\mathrm{GeV}$ for $V = W, Z$, the hadronic decay products of an EW boson are no longer contained within a single $R = 0.8$ jet and the boson tag loses its meaning. At $5\,\mathrm{TeV}$ the same decay topology scales up: the soft boson is now well above the threshold and survives the selection, so all three jets are available and the full three-body mass is reconstructed, giving a visible peak despite the smaller production cross-section.

\begin{figure}[htbp!]
    \centering
    \begin{tabular}{@{}cc@{}}
        \includegraphics[width=0.48\textwidth]{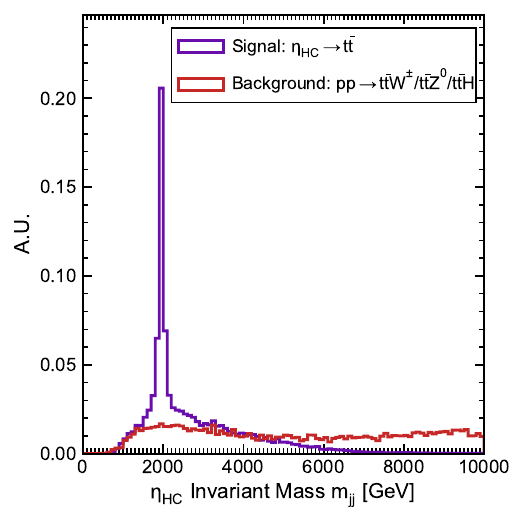} &
        \includegraphics[width=0.48\textwidth]{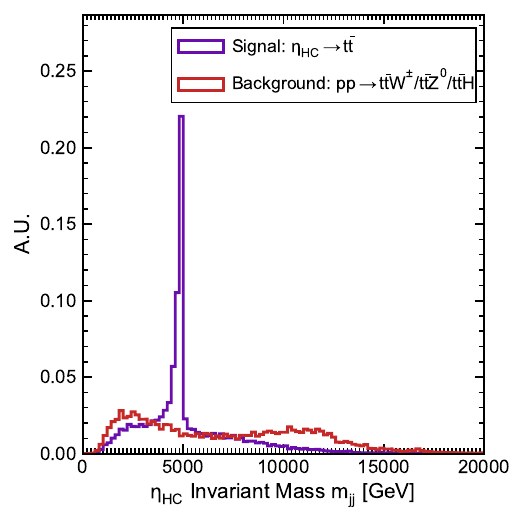} \\
    \end{tabular}
    \caption{Dijet invariant mass distribution for the $\etaT$ signal
    (purple) and the associated top-pair background (red), at
    $M_{\etaT} = 2$~TeV (left) and $5$~TeV (right), generated with $\sqrt{\hat{s}}_{\min} = 10$~TeV and $12$~TeV respectively.}
    \label{fig:eta_mass}
\end{figure}
\paragraph{Pseudoscalar $\etaT$ search.}
The pseudoscalar $\etaT$, with a mass of $2\,(5)\,\mathrm{TeV}$ for Benchmark~1~(2), is produced with $p_T$ comparable to its mass. Its dominant decay is into a top pair, $\etaT \to t\bar{t}$. Since the resonance is boosted, the two tops share its momentum asymmetrically, one carrying approximately $2.5$ times the $p_T$ of the other, and they remain angularly separated with $\Delta R \sim 2$, so each top gives its own resolved jet rather than a single merged jet. Each top is itself strongly boosted, so that its three-prong hadronic decay remains collimated within the jet radius and merges into a single large-$R$ jet whose groomed mass lies close to the top mass. Top candidates are identified by a groomed mass in the range $150$--$350\,\mathrm{GeV}$, and the $\etaT$ invariant mass is reconstructed from the leading pair of top-tagged jets,
\begin{equation}
  m_{\etaT}^{\mathrm{reco}} = \sqrt{\left(p_{j_0}+p_{j_1}\right)^{2}} ,
  \label{eq:meta_reco}
\end{equation}
where $j_0$ and $j_1$ are the leading and subleading top-tagged jets in $p_T$, in events with a jet multiplicity $n \geq 3$ and at least two top-tagged jets. The reconstructed spectrum, shown in Figure~\ref{fig:eta_mass}, exhibits a peak at the generated mass on top of a broad continuum, arising from the background, the associated production of top pairs, $pp \to t\bar{t}\,W^{\pm}/Z/H$, which is reconstructed in the same way as the signal.

\FloatBarrier

\subsection{Signal significance}
\label{sec:pheno:significance}

To quantify the significance of the signal at the FCC-hh, we employ two complementary strategies: first, we select events containing one of the heavy resonances, $\mathcal{R} = \rhoT,\ \omT,\ \etaT$; second, we consider the inclusive production of multi-EW bosons. For the resonance searches, we bin the events over the appropriate reconstructed invariant mass, while for the inclusive EW boson one over the $p_T$ distribution.
The signal sensitivity of each search channel is quantified using the
Asimov significance summed over the individual bins~\cite{Cowan:2010js},
\begin{equation}
  \mathcal{Z}
  = \sqrt{\,\sum_i 2\!\left[
      (s_i+b_i)\ln\!\left(1+\frac{s_i}{b_i}\right)-s_i
    \right]}\, ,
  \label{eq:zsig}
\end{equation}
where $s_i$ ($b_i$) denote the expected signal (background) yields
in bin $i$, evaluated for an integrated luminosity of
$\mathcal{L} = 30\,\mathrm{ab}^{-1}$. 
The backgrounds are the SM multi-boson and
top-associated processes introduced above.
To obtain the yields in the individual bins, we estimate the effective  cross-section for the resonances $\rhoT$, $\omT$, and $\etaT$, or the inclusive EW bosons, as
\begin{equation}
  \sigma^{\prime}_{\rm sig}
  \;=\; \sigma_{\rm sig}\,\epsilon_{\rm gen}\,\mathcal{B}\,X_{\rm Br}\,,
  \qquad
  \sigma^{\prime}_{\rm bkg}
  \;=\; \sigma_{\rm bkg}\,X_{\rm Br}\,,
  \label{eq:sigmaeff}
\end{equation}
where
$\epsilon_{\rm gen}$ is the generator-level fraction of events containing the resonance and $\mathcal{B}$ is the branching ratio of the resonance into the
final state under study ($\epsilon_{\rm gen} = \mathcal{B}=1$ for the inclusive channel); furthermore $\sigma_{\rm sig}$ is the leading-order cross-section for the process $pp\rightarrow q_{\rm HC}\bar q_{\rm HC}$, $\sigma_{\rm bkg}$ the corresponding one for the backgrounds.
Throughout the analysis, only the hadronic decays of the EW bosons are
reconstructed, so both the signal and the background cross-sections carry the
additional factor $X_{\rm Br}$ of the hadronic branching fraction of the
final state. Each entry of a reconstructed mass spectrum is a combination of
$n_{\rm min}$ boson-tagged jets,  so that $X_{\rm Br} = \varepsilon^{\,n_{\rm min}}$ with
$\varepsilon \simeq 0.68$ the hadronic branching fraction per boson averaged
over the flavour composition of the samples, built from
$\mathcal{B}(W\to q\bar q^{\,\prime}) = 0.674$ and
$\mathcal{B}(Z\to q\bar q) = 0.699$. Here $n_{\rm min} = 2$ for the di-boson
decays of the $\rhoT$, $n_{\rm min} = 3$ for the $\omT$, and $n_{\rm min} = 2$
for the two $W$ bosons from the top decays of the $\etaT$. The same factor applies to
the signal and to the background, since both spectra are built by the same
combinatorial procedure, so it reduces the yields of each channel while leaving
$S/B$ unchanged. 

The values of the parameters and of the significances we obtained are listed in
Tables~\ref{tab:tc_significance_bp1} and~\ref{tab:tc_significance_bp2} for the two benchmarks.
The $\omT$ channel
requires a separate normalization because its trijet reconstruction is
performed only after the jets already assigned to a $\rhoT$ candidate have
been vetoed. Consequently, only the reduced subset of events surviving this
veto enters the $\omT$ search. 
The same procedure is applied to the background, using its corresponding cross-sections and event counts. For Benchmark 1, we provide two complementary significances: the local significance is evaluated in a window centred on the true resonance mass,
$1800$--$2200\,\mathrm{GeV}$, in $50\,\mathrm{GeV}$ bins, and quantifies the strength of the excess once
the resonance position is specified; the global significance additionally
accounts for the look-elsewhere effect~\cite{Gross:2010qma} over the full accessible range,
$500$--$12000\,\mathrm{GeV}$. Both results, together with the
signal-to-background ratio in the local window, are given in
Table~\ref{tab:tc_significance_bp1}. For Benchmark 2, for which the resonances
are centred at $5\,\mathrm{TeV}$, the corresponding window is
$4500$--$5500\,\mathrm{GeV}$, in $100\,\mathrm{GeV}$ bins, with the results summarized in
Table~\ref{tab:tc_significance_bp2}.

\begin{table}[htbp]
\centering
\small
\caption{\textbf{Benchmark 1}: Signal significance for the three QCD-like
resonance searches at $2\,\mathrm{TeV}$. No global significance is quoted for
the $\omT$, whose local excess does not survive the look-elsewhere
correction.}
\label{tab:tc_significance_bp1}
\renewcommand{\arraystretch}{1.3}
\begin{tabular}{lccccccc}
\hline\hline
\textbf{Resonance} & \textbf{Channel} &
\textbf{Gen.\ fraction} & $\mathcal{B}\,(\%)$ &
$X_{\rm Br}$ &
$S/B$ &
$\mathcal{Z}_\mathrm{local}$ &
$\mathcal{Z}_\mathrm{global}$ \\
\hline
$\rhoT$ & $W^+W^-/W^\pm Z$ & $0.146$ & $27$ & $0.47$
  & $0.214$ & $5.32\,\sigma$ & $4.95\,\sigma$ \\[2pt]
$\omT$  & $W^+ZW^-$         & $0.0715$  & $100$ & $0.31$
  & $0.080$ & $1.66\,\sigma$ &  \\[2pt]
$\etaT$ & $t\bar{t}$        & $0.151$ & $100$ & $0.46$
  & $0.144$ & $3.71\,\sigma$ & $2.23\,\sigma$ \\
\hline\hline
\end{tabular}
\end{table}
\begin{table}[htbp]
\centering
\small
\caption{\textbf{Benchmark 2}: Signal significance for the three QCD-like
resonance searches at $5\,\mathrm{TeV}$.}

\label{tab:tc_significance_bp2}
\renewcommand{\arraystretch}{1.3}
\begin{tabular}{lcccccc}
\hline\hline
\textbf{Resonance} & \textbf{Channel} &
\textbf{Gen.\ fraction} & $\mathcal{B}\,(\%)$ &
$X_{\rm Br}$ &
$S/B$ &
$\mathcal{Z}$ \\
\hline
$\rhoT$ & $W^+W^-/W^\pm Z$ & $0.0939$ & $60$ & $0.46$
  & $0.133$ & $2.23\,\sigma$ \\[2pt]
$\omT$  & $W^+ZW^-$         & $0.0292$  & $100$ & $0.31$
  & $0.020$ & $0.69\,\sigma$  \\[2pt]
$\etaT$ & $t\bar{t}$        & $0.0523$ & $100$ & $0.46$
  & $0.018$ & $0.36\,\sigma$ \\
\hline\hline
\end{tabular}
\end{table}

For the inclusive EW-boson channel, the bins are defined in the $p_T$ distribution of
all EW-boson-tagged jets, with a bin width of $20\,\mathrm{GeV}$ over the
range $0$--$10\,\mathrm{TeV}$ for both benchmarks. The corresponding
leading-jet $p_T$ distribution is shown in Figure~\ref{jetpT}. The EW bosons
produced during hyperquark fragmentation and hadronization, as well as those
arising from the decays of the vector resonances, are predominantly hard and
central, whereas the SM bosons produced in the background processes are
predominantly softer and more forward. The signal therefore produces a
broad excess across the $p_T$ spectrum rather than a localized resonance
peak. The resulting inclusive significance consequently exceeds the
conventional $5\sigma$ discovery threshold over the background fluctuations.

\subsection{Discussion of the collider analysis}
\label{sec:Z_discussion}

The discovery potential of the QCD-like resonances is explored in the TC limit, with Benchmark~1 providing the primary reference point. As shown in Table~\ref{tab:tc_significance_bp1}, at a mass of $2\,\mathrm{TeV}$ the techni-rho $\rhoT$ and the techni-eta-prime $\etaT$ achieve local significances of $5.3\sigma$ and $3.7\sigma$ respectively, corresponding to discovery- and evidence-level sensitivity, so that these channels would be within reach of the FCC-hh. The techni-omega $\omT$ reaches only about $1.7\sigma$ at the same resonance mass. Its reduced sensitivity originates in the reconstruction of a three-body final state, in which all three EW bosons from the $\omT$ decay must be separately resolved and tagged. Because the three bosons have an asymmetric $p_T$ distribution, the softest one carries only a few hundred GeV and frequently falls below the jet $p_T^{\min}$ selection. This reduces the reconstruction efficiency relative to the $\rhoT$. A smaller generator-level production fraction further suppresses the effective cross-section in Eq.~\eqref{eq:sigmaeff}. In addition, the veto applied to jets already used in the $\rhoT$ reconstruction removes a fraction of the remaining $\omT$ candidates, although it suppresses proportionally more combinatorial background. The latter effect is a consequence of performing the two searches sequentially on the same jet collection, and should therefore not be interpreted as an intrinsic limitation of the $\omT$ signal.

The existing HL-LHC constraints on a QCD-like $\rhoT$ resonance of $2\,\mathrm{TeV}$ motivate exploring larger resonance masses, corresponding
to higher $\LambdaT$ and moving away from the technicolor limit. Since
the hyperquark vector currents are independent of the misalignment angle,
the resonance production can be studied in this extended region of the
parameter space. As summarized in Table~\ref{tab:tc_significance_bp2}, for
Benchmark 2, where the resonance masses are at $5\,\mathrm{TeV}$, $\rhoT$ reaches at most about $2\sigma$,
for a branching ratio of $60\%$ at a heavy-pion mass of $2\,\mathrm{TeV}$.
The corresponding $\omT$ and $\etaT$ signals remain below $1\sigma$.
The reduction in significance at higher resonance masses has a kinematic
origin. The multi-EW-boson, spin-1, and spin-0 resonance spectrum is produced
through the fragmentation and hadronization of the hyperquark pair, and the
resulting multiplicity depends on the available showering phase space. This
phase space is determined by the energy available above the hyperquark
pair-production threshold,
\begin{equation}
\Delta
= \sqrt{\hat{s}}-2m_{q_{\rm HC}},
\end{equation}
which is conveniently measured relative to $\Lambda_{\rm HC}$. Increasing
$\Lambda_{\rm HC}$ reduces the available phase space for showering at a fixed
$\sqrt{\hat{s}}$, leading to a shorter shower and consequently fewer
EW bosons and resonances in the final state. Increasing
$\sqrt{\hat{s}}_{\min}$ can partially compensate for this effect by providing
additional phase space for showering and thereby restoring a larger
multiplicity of composite states.
However, raising $\sqrt{\hat{s}}_{\min}$ also reduces the production rate. The signal leading-order cross-section decreases rapidly as the required $\sqrt{\hat{s}}_{\min}$  is increased, while the corresponding SM multi-boson and $t\bar t$ backgrounds fall less rapidly over the relevant region. As a result, the background becomes increasingly important in the resonance windows, leading to the reduced significances observed for Benchmark 2.

\FloatBarrier
\section{Conclusions}
\label{sec:conclusion}

The search for heavy resonances in the bosonic final state, both through direct and associated production, is one of the leading agendas at the LHC and its high-luminosity upgrade, HL-LHC, as well as the proposed future circular hadronic collider FCC-hh. 
In particular, the ongoing searches at CMS and ATLAS 
can probe masses up to 3--5 TeV through DY and VBF channels by the end of the HL phase.
While these searches have considerable potential to probe different parts of the model parameter space, they are not exclusive to composite models and hence do not unveil the underlying origins. 

This steers our attention to  their production by means of the hadronization of constituent fermions, hyperquarks, produced in the process $pp\rightarrow q_{\rm HC}\bar q_{\rm HC}$ \cite{Cacciapaglia:2025multiboson}. The process necessitates that the partonic centre-of-mass energy satisfies $\sqrt{\hat{s}}_{\min} > \Lambda_{\rm HC}$, thereby making the FCC-hh one of the ideal candidates and the focus of attention in this paper.
This process leads to the simultaneous formation of EW bosons as well as heavy composite resonances specific to the coset of the composite Higgs model. The eventual decay of these resonances, particularly in the bosonic mode, leads to a multi-boson splash at the detector.  Working with a model based on the coset
$(\SU{4}_L \times \SU{4}_R) / \SU{4}_V$, we  developed the spectroscopy of the model that can be applied both to the technicolor-like limit ($f = v$) and to the composite Higgs one: fifteen techni-rho states $\rhoT^a$ together with
the techni-omega $\omT$ and the techni-eta-prime $\etaT$, and characterized their decays. Among them,  $\rhoT^{\pm,0}$, $\omT$ and  $\etaT$ are particularly interesting and their collider reach is presented in this paper. The study of   $\rhoT^{\pm,0}$ offers a
genuinely model-independent handle on the compositeness of the
EW sector regardless of where the true vacuum resides.  The techni-omega $\omT$ decays instead through the Wess--Zumino--Witten
anomaly into three EW bosons, in exact analogy with $\omega \to
\pi^+\pi^-\pi^0$ in QCD, while the techni-eta-prime $\etaT$ decays
dominantly to a top-antitop pair through extended technicolor
interactions, its mass carrying an explicit dependence on the number of
hypercolors that distinguishes it from the confinement-scale masses of
$\rhoT^{\pm,0}$ and $\omT$. While each of these resonances is being probed individually, a common origin for their production by means of hadronization offers not only a complementary test but also an insight into their composite origins. 

This analysis paves the way for not only investigating the complete spectroscopy of the model we present but also going beyond  the QCD-like structure of the model: a confining group whose fundamental representation is complex and the eventual formation of composite states by hadronization.  Other constructions 
 built on pseudo-real hypercolor groups such
as $\Sp(2N)$, exemplified by the minimal $\SU{4}/\Sp(4)$
coset~\cite{Cacciapaglia:2014uja,Franzosi:2016aoo,Bennett:2017kga}, or on
real representations leading to cosets such as
$\SU{5}/\mathrm{SO}(5)$~\cite{Agugliaro:2019wtf,Ryttov:2008xe}, do not
admit the same treatment and instead require dedicated fragmentation and
hadronization modelling tailored to their distinct group-theoretic
structure~\cite{Kosower:1984aw}.
Extending the event-generator tools used here to these non-$\SU{N}$
hypercolor groups remains an open and necessary step toward a full general collider phenomenology of composite EW symmetry breaking.

\acknowledgments
We thank Stephen Mrenna for his valuable assistance in implementing the decay of the resonance $\etaT$ into a top-antitop quark pair within the \textsc{Pythia}~8.312 Hidden Valley framework. S.~K. is supported by the FWF project number P 36947-N. The research of A.~K.~S. is supported by the Institute Fellowship from the Indian Institute of Technology Delhi, India.

\begin{appendix}
\section{Generator Basis, Vacuum Structure, and Commutation Relations for $(\SU{4}_L\times\SU{4}_R)/\SU{4}_V$}
\label{app:generators}

\subsection*{A.1\quad $\SU{4}$ generators and the reference vacuum}

The global symmetry group $\mathcal{G} = \SU{4}_L\times\SU{4}_R$ has $d_\mathcal{G} = 30$ generators
$T^A$, which decompose into the fifteen unbroken generators of $\mathcal{H} = \SU{4}_V$ and
the fifteen broken generators of the coset $\mathcal{G}/\mathcal{H}$,
\begin{equation}
  T^A = \bigl\{\mathcal{T}^a,\;\mathcal{X}^a\bigr\}.
  \label{eq:gendecomp_app}
\end{equation}
In the $8$-dimensional representation acting on the column vector $(\qL,\qR)^T$,
all generators take the block form~\cite{Preskill:1980mz}
\begin{equation}
  T^A =
  \begin{pmatrix}
    T^A_L & 0 \\
    0 & -(T^A_R)^T
  \end{pmatrix},
  \label{eq:Tblock8_app}
\end{equation}
and the reference vacuum $\Sigma_0$, which preserves the EW symmetry at
$\theta = 0$, is~\cite{Ma:2015gra,Agugliaro:2019wtf}
\begin{equation}
  \Sigma_0 =
  \begin{pmatrix}
    0 & \mathbb{1}_{4\times4} \\
    \mathbb{1}_{4\times4} & 0
  \end{pmatrix}.
  \label{eq:refvac_app}
\end{equation}

\subsection*{A.2\quad Unbroken and broken generators at the reference vacuum}

A generator is unbroken if it leaves $\Sigma_0$ invariant~\cite{Cacciapaglia:2014uja},
\begin{align}
  \mathcal{T}^a\,\Sigma_0 + \Sigma_0\,(\mathcal{T}^a)^T &= 0
  \quad \text{(unbroken)},
  \label{eq:unbrok_app}
  \\[4pt]
  \mathcal{X}^a\,\Sigma_0 - \Sigma_0\,(\mathcal{X}^a)^T &= 0
  \quad \text{(broken)}.
  \label{eq:brok_app}
\end{align}
Substituting the block form~\eqref{eq:Tblock8_app} gives $T_L = T_R$ for the
unbroken generators and $T_L = -T_R$ for the broken ones, so the two sets take
the explicit $8\times8$ forms
\begin{equation}
  \mathcal{T}^a =
  \begin{pmatrix}
    \Lambda^a & 0 \\
    0 & -(\Lambda^a)^T
  \end{pmatrix},
  \qquad
  \mathcal{X}^a =
  \begin{pmatrix}
    \Lambda^a & 0 \\
    0 & (\Lambda^a)^T
  \end{pmatrix},
  \label{eq:unbroken8_ref_app}
\end{equation}
where $\Lambda^a$ are the $4\times4$ $\SU{4}$ generators in the fundamental
representation normalized as $\tr[\Lambda^a\Lambda^b] = \half\delta^{ab}$.

\subsection*{A.3\quad Vacuum alignment and the misaligned generators}

The true vacuum, in which the EW symmetry is broken by an angle $\theta$,
is obtained by rotating the reference vacuum with the matrix~\cite{Ma:2015gra}
\begin{equation}
  \Omega_4 =
  \begin{pmatrix}
    \cos\tfrac{\theta}{2}\,I_2 & \sin\tfrac{\theta}{2}\,I_2 \\[4pt]
    -\sin\tfrac{\theta}{2}\,I_2 & \cos\tfrac{\theta}{2}\,I_2
  \end{pmatrix},
  \label{eq:Omega4_app}
\end{equation}
so that the true vacuum in the $4\times4$ flavour space and in the full $8\times8$
embedding are
\begin{equation}
  \Sigma^{(4)}_0(\theta) = \Omega^{2}_4,
  \qquad
  \Sigma_0 =
  \begin{pmatrix}
    0 & \Sigma^{(4)}_0(\theta) \\
    \bigl(\Sigma^{(4)}_0(\theta)\bigr)^T & 0
  \end{pmatrix}.
  \label{eq:Sigmatilde_app}
\end{equation}
Imposing the unbroken condition~\eqref{eq:unbrok_app} with $\Sigma_0$ for $\theta \neq 0$ and substituting the block form gives
\begin{equation}
  T_L\,\Sigma^{(4)}_0(\theta) \;-\; \Sigma^{(4)}_0(\theta)\,T_R \;=\; 0.
  \label{eq:vacalign1_app}
\end{equation}
Using $\Sigma^{(4)}_0(\theta) = \Omega_4\,\Omega_4$ this becomes
\begin{equation}
  T_L\,\Omega_4\,\Omega_4 \;-\; \Omega_4\,\Omega_4\,T_R \;=\; 0
  \quad\Longrightarrow\quad
  \Omega_4^\dagger\,T_L\,\Omega_4 \;=\; \Omega_4\,T_R\,\Omega_4^\dagger.
  \label{eq:vacalign_app}
\end{equation}
The unbroken condition in the true vacuum therefore reduces to
$\widetilde{T}_L = \widetilde{T}_R$, where the misaligned left and right components
are defined by
\begin{equation}
  \widetilde{T}_L \;\equiv\; \Omega_4^\dagger\,T_L\,\Omega_4 \;\equiv\; \widetilde{\Lambda}^a,
  \qquad
  \widetilde{T}_R \;\equiv\; \Omega_4\,T_R\,\Omega_4^\dagger \;\equiv\; \widetilde{\Lambda}^a,
  \label{eq:tildedef_app}
\end{equation}
so both sides equal the same rotated $4\times4$ matrix $\widetilde{\Lambda}^a$.
The $8\times8$ misaligned unbroken generators are then
\begin{equation}
  \widetilde{\mathcal{T}}^a \;=\;
  \begin{pmatrix}
    \widetilde{\Lambda}^a & 0 \\[2pt]
    0 & -(\widetilde{\Lambda}^a)^T
  \end{pmatrix},
  \label{eq:tildeTgen_app}
\end{equation}
and using the same procedure for the broken condition~\eqref{eq:brok_app},
\begin{equation}
  \widetilde{\mathcal{X}}^a \;=\;
  \begin{pmatrix}
    \widetilde{\Lambda}^a & 0 \\[2pt]
    0 &  (\widetilde{\Lambda}^a)^T
  \end{pmatrix}.
  \label{eq:tildeXgen_app}
\end{equation}
The $8\times8$ forms above are used to define the vacuum and to separate the
broken from the unbroken directions, Eqs.~\eqref{eq:unbrok_app}
and~\eqref{eq:brok_app}. They are normalized as
$\tr[\widetilde{\mathcal{T}}^a\widetilde{\mathcal{T}}^b]
=\tr[\widetilde{\mathcal{X}}^a\widetilde{\mathcal{X}}^b]=\delta^{ab}$, and since both
are built from the same $4\times4$ block $\widetilde{\Lambda}^a$, a trace in the
$8\times8$ form is twice the corresponding $4\times4$ one. It is this block,
normalized as
$\tr[\widetilde{\Lambda}^a\widetilde{\Lambda}^b] = \tfrac{1}{2}\delta^{ab}$, that
enters the current formula~\eqref{eq:current_formula} and the hidden local symmetry
construction of Section~\ref{sec:HLS} and Appendix~\ref{app:HLS}.
At $\theta = 0$, $\widetilde{\Lambda}^a = \Lambda^a$ and the unrotated forms
$\mathcal{T}^a$, $\mathcal{X}^a$ of A.2 are recovered.
For generators $\mathcal{T}^a$ that commute with $\Omega_4$, the rotation
in~\eqref{eq:tildedef_app} leaves them unchanged for all $\theta$, making them
$\theta$-independent.

\subsection*{A.4\quad The fifteen misaligned generators of $\SU{4}_V$}

The explicit $8\times8$ misaligned generators are obtained by substituting
$\Omega_4$ into Eq.~\eqref{eq:tildedef_app}.
We write $c \equiv \cos\theta$, $s \equiv \sin\theta$,
$c_{1/2} \equiv \cos(\theta/2)$, $s_{1/2} \equiv \sin(\theta/2)$.

\begin{equation}
\widetilde{\mathcal{T}}^1 =
\begin{pmatrix}
0 & \tfrac{1}{2}c_{1/2}^2 & 0 & -\tfrac{s}{4} & 0 & 0 & 0 & 0 \\
\tfrac{1}{2}c_{1/2}^2 & 0 & -\tfrac{s}{4} & 0 & 0 & 0 & 0 & 0 \\
0 & -\tfrac{s}{4} & 0 & \tfrac{1}{2}s_{1/2}^2 & 0 & 0 & 0 & 0 \\
-\tfrac{s}{4} & 0 & \tfrac{1}{2}s_{1/2}^2 & 0 & 0 & 0 & 0 & 0 \\
0 & 0 & 0 & 0 & 0 & -\tfrac{1}{2}c_{1/2}^2 & 0 & \tfrac{s}{4} \\
0 & 0 & 0 & 0 & -\tfrac{1}{2}c_{1/2}^2 & 0 & \tfrac{s}{4} & 0 \\
0 & 0 & 0 & 0 & 0 & \tfrac{s}{4} & 0 & -\tfrac{1}{2}s_{1/2}^2 \\
0 & 0 & 0 & 0 & \tfrac{s}{4} & 0 & -\tfrac{1}{2}s_{1/2}^2 & 0
\end{pmatrix}
\label{eq:Ttilde1_app}
\end{equation}

\begin{equation}
\widetilde{\mathcal{T}}^2 =
\begin{pmatrix}
0 & -\tfrac{i}{2}c_{1/2}^2 & 0 & \tfrac{is}{4} & 0 & 0 & 0 & 0 \\
\tfrac{i}{2}c_{1/2}^2 & 0 & -\tfrac{is}{4} & 0 & 0 & 0 & 0 & 0 \\
0 & \tfrac{is}{4} & 0 & -\tfrac{i}{2}s_{1/2}^2 & 0 & 0 & 0 & 0 \\
-\tfrac{is}{4} & 0 & \tfrac{i}{2}s_{1/2}^2 & 0 & 0 & 0 & 0 & 0 \\
0 & 0 & 0 & 0 & 0 & -\tfrac{i}{2}c_{1/2}^2 & 0 & \tfrac{is}{4} \\
0 & 0 & 0 & 0 & \tfrac{i}{2}c_{1/2}^2 & 0 & -\tfrac{is}{4} & 0 \\
0 & 0 & 0 & 0 & 0 & \tfrac{is}{4} & 0 & -\tfrac{i}{2}s_{1/2}^2 \\
0 & 0 & 0 & 0 & -\tfrac{is}{4} & 0 & \tfrac{i}{2}s_{1/2}^2 & 0
\end{pmatrix}
\label{eq:Ttilde2_app}
\end{equation}

\begin{equation}
\widetilde{\mathcal{T}}^3 =
\begin{pmatrix}
\tfrac{1}{2}c_{1/2}^2 & 0 & -\tfrac{s}{4} & 0 & 0 & 0 & 0 & 0 \\
0 & -\tfrac{1}{2}c_{1/2}^2 & 0 & \tfrac{s}{4} & 0 & 0 & 0 & 0 \\
-\tfrac{s}{4} & 0 & \tfrac{1}{2}s_{1/2}^2 & 0 & 0 & 0 & 0 & 0 \\
0 & \tfrac{s}{4} & 0 & -\tfrac{1}{2}s_{1/2}^2 & 0 & 0 & 0 & 0 \\
0 & 0 & 0 & 0 & -\tfrac{1}{2}c_{1/2}^2 & 0 & \tfrac{s}{4} & 0 \\
0 & 0 & 0 & 0 & 0 & \tfrac{1}{2}c_{1/2}^2 & 0 & -\tfrac{s}{4} \\
0 & 0 & 0 & 0 & \tfrac{s}{4} & 0 & -\tfrac{1}{2}s_{1/2}^2 & 0 \\
0 & 0 & 0 & 0 & 0 & -\tfrac{s}{4} & 0 & \tfrac{1}{2}s_{1/2}^2
\end{pmatrix}
\label{eq:Ttilde3_app}
\end{equation}

\begin{equation}
\widetilde{\mathcal{T}}^4 =
\begin{pmatrix}
0 & \tfrac{1}{2}s_{1/2}^2 & 0 & \tfrac{s}{4} & 0 & 0 & 0 & 0 \\
\tfrac{1}{2}s_{1/2}^2 & 0 & \tfrac{s}{4} & 0 & 0 & 0 & 0 & 0 \\
0 & \tfrac{s}{4} & 0 & \tfrac{1}{2}c_{1/2}^2 & 0 & 0 & 0 & 0 \\
\tfrac{s}{4} & 0 & \tfrac{1}{2}c_{1/2}^2 & 0 & 0 & 0 & 0 & 0 \\
0 & 0 & 0 & 0 & 0 & -\tfrac{1}{2}s_{1/2}^2 & 0 & -\tfrac{s}{4} \\
0 & 0 & 0 & 0 & -\tfrac{1}{2}s_{1/2}^2 & 0 & -\tfrac{s}{4} & 0 \\
0 & 0 & 0 & 0 & 0 & -\tfrac{s}{4} & 0 & -\tfrac{1}{2}c_{1/2}^2 \\
0 & 0 & 0 & 0 & -\tfrac{s}{4} & 0 & -\tfrac{1}{2}c_{1/2}^2 & 0
\end{pmatrix}
\label{eq:Ttilde4_app}
\end{equation}

\begin{equation}
\widetilde{\mathcal{T}}^5 =
\begin{pmatrix}
0 & -\tfrac{i}{2}s_{1/2}^2 & 0 & -\tfrac{is}{4} & 0 & 0 & 0 & 0 \\
\tfrac{i}{2}s_{1/2}^2 & 0 & \tfrac{is}{4} & 0 & 0 & 0 & 0 & 0 \\
0 & -\tfrac{is}{4} & 0 & -\tfrac{i}{2}c_{1/2}^2 & 0 & 0 & 0 & 0 \\
\tfrac{is}{4} & 0 & \tfrac{i}{2}c_{1/2}^2 & 0 & 0 & 0 & 0 & 0 \\
0 & 0 & 0 & 0 & 0 & -\tfrac{i}{2}s_{1/2}^2 & 0 & -\tfrac{is}{4} \\
0 & 0 & 0 & 0 & \tfrac{i}{2}s_{1/2}^2 & 0 & \tfrac{is}{4} & 0 \\
0 & 0 & 0 & 0 & 0 & -\tfrac{is}{4} & 0 & -\tfrac{i}{2}c_{1/2}^2 \\
0 & 0 & 0 & 0 & \tfrac{is}{4} & 0 & \tfrac{i}{2}c_{1/2}^2 & 0
\end{pmatrix}
\label{eq:Ttilde5_app}
\end{equation}

\begin{equation}
\widetilde{\mathcal{T}}^6 =
\begin{pmatrix}
\tfrac{1}{2}s_{1/2}^2 & 0 & \tfrac{s}{4} & 0 & 0 & 0 & 0 & 0 \\
0 & -\tfrac{1}{2}s_{1/2}^2 & 0 & -\tfrac{s}{4} & 0 & 0 & 0 & 0 \\
\tfrac{s}{4} & 0 & \tfrac{1}{2}c_{1/2}^2 & 0 & 0 & 0 & 0 & 0 \\
0 & -\tfrac{s}{4} & 0 & -\tfrac{1}{2}c_{1/2}^2 & 0 & 0 & 0 & 0 \\
0 & 0 & 0 & 0 & -\tfrac{1}{2}s_{1/2}^2 & 0 & -\tfrac{s}{4} & 0 \\
0 & 0 & 0 & 0 & 0 & \tfrac{1}{2}s_{1/2}^2 & 0 & \tfrac{s}{4} \\
0 & 0 & 0 & 0 & -\tfrac{s}{4} & 0 & -\tfrac{1}{2}c_{1/2}^2 & 0 \\
0 & 0 & 0 & 0 & 0 & \tfrac{s}{4} & 0 & \tfrac{1}{2}c_{1/2}^2
\end{pmatrix}
\label{eq:Ttilde6_app}
\end{equation}

\begin{equation}
\widetilde{\mathcal{T}}^7 =
\begin{pmatrix}
0 & \tfrac{s}{2\sqrt{2}} & 0 & \tfrac{c}{2\sqrt{2}} & 0 & 0 & 0 & 0 \\
\tfrac{s}{2\sqrt{2}} & 0 & \tfrac{c}{2\sqrt{2}} & 0 & 0 & 0 & 0 & 0 \\
0 & \tfrac{c}{2\sqrt{2}} & 0 & -\tfrac{s}{2\sqrt{2}} & 0 & 0 & 0 & 0 \\
\tfrac{c}{2\sqrt{2}} & 0 & -\tfrac{s}{2\sqrt{2}} & 0 & 0 & 0 & 0 & 0 \\
0 & 0 & 0 & 0 & 0 & -\tfrac{s}{2\sqrt{2}} & 0 & -\tfrac{c}{2\sqrt{2}} \\
0 & 0 & 0 & 0 & -\tfrac{s}{2\sqrt{2}} & 0 & -\tfrac{c}{2\sqrt{2}} & 0 \\
0 & 0 & 0 & 0 & 0 & -\tfrac{c}{2\sqrt{2}} & 0 & \tfrac{s}{2\sqrt{2}} \\
0 & 0 & 0 & 0 & -\tfrac{c}{2\sqrt{2}} & 0 & \tfrac{s}{2\sqrt{2}} & 0
\end{pmatrix}
\label{eq:Ttilde7_app}
\end{equation}

\begin{equation}
\widetilde{\mathcal{T}}^8 =
\begin{pmatrix}
0 & -\tfrac{is}{2\sqrt{2}} & 0 & -\tfrac{ic}{2\sqrt{2}} & 0 & 0 & 0 & 0 \\
\tfrac{is}{2\sqrt{2}} & 0 & \tfrac{ic}{2\sqrt{2}} & 0 & 0 & 0 & 0 & 0 \\
0 & -\tfrac{ic}{2\sqrt{2}} & 0 & \tfrac{is}{2\sqrt{2}} & 0 & 0 & 0 & 0 \\
\tfrac{ic}{2\sqrt{2}} & 0 & -\tfrac{is}{2\sqrt{2}} & 0 & 0 & 0 & 0 & 0 \\
0 & 0 & 0 & 0 & 0 & -\tfrac{is}{2\sqrt{2}} & 0 & -\tfrac{ic}{2\sqrt{2}} \\
0 & 0 & 0 & 0 & \tfrac{is}{2\sqrt{2}} & 0 & \tfrac{ic}{2\sqrt{2}} & 0 \\
0 & 0 & 0 & 0 & 0 & -\tfrac{ic}{2\sqrt{2}} & 0 & \tfrac{is}{2\sqrt{2}} \\
0 & 0 & 0 & 0 & \tfrac{ic}{2\sqrt{2}} & 0 & -\tfrac{is}{2\sqrt{2}} & 0
\end{pmatrix}
\label{eq:Ttilde8_app}
\end{equation}

\begin{equation}
\widetilde{\mathcal{T}}^9 =
\begin{pmatrix}
\tfrac{s}{2\sqrt{2}} & 0 & \tfrac{c}{2\sqrt{2}} & 0 & 0 & 0 & 0 & 0 \\
0 & -\tfrac{s}{2\sqrt{2}} & 0 & -\tfrac{c}{2\sqrt{2}} & 0 & 0 & 0 & 0 \\
\tfrac{c}{2\sqrt{2}} & 0 & -\tfrac{s}{2\sqrt{2}} & 0 & 0 & 0 & 0 & 0 \\
0 & -\tfrac{c}{2\sqrt{2}} & 0 & \tfrac{s}{2\sqrt{2}} & 0 & 0 & 0 & 0 \\
0 & 0 & 0 & 0 & -\tfrac{s}{2\sqrt{2}} & 0 & -\tfrac{c}{2\sqrt{2}} & 0 \\
0 & 0 & 0 & 0 & 0 & \tfrac{s}{2\sqrt{2}} & 0 & \tfrac{c}{2\sqrt{2}} \\
0 & 0 & 0 & 0 & -\tfrac{c}{2\sqrt{2}} & 0 & \tfrac{s}{2\sqrt{2}} & 0 \\
0 & 0 & 0 & 0 & 0 & \tfrac{c}{2\sqrt{2}} & 0 & -\tfrac{s}{2\sqrt{2}}
\end{pmatrix}
\label{eq:Ttilde9_app}
\end{equation}

\begin{equation}
\widetilde{\mathcal{T}}^{10} =
\begin{pmatrix}
\tfrac{s}{2\sqrt{2}} & 0 & \tfrac{c}{2\sqrt{2}} & 0 & 0 & 0 & 0 & 0 \\
0 & \tfrac{s}{2\sqrt{2}} & 0 & \tfrac{c}{2\sqrt{2}} & 0 & 0 & 0 & 0 \\
\tfrac{c}{2\sqrt{2}} & 0 & -\tfrac{s}{2\sqrt{2}} & 0 & 0 & 0 & 0 & 0 \\
0 & \tfrac{c}{2\sqrt{2}} & 0 & -\tfrac{s}{2\sqrt{2}} & 0 & 0 & 0 & 0 \\
0 & 0 & 0 & 0 & -\tfrac{s}{2\sqrt{2}} & 0 & -\tfrac{c}{2\sqrt{2}} & 0 \\
0 & 0 & 0 & 0 & 0 & -\tfrac{s}{2\sqrt{2}} & 0 & -\tfrac{c}{2\sqrt{2}} \\
0 & 0 & 0 & 0 & -\tfrac{c}{2\sqrt{2}} & 0 & \tfrac{s}{2\sqrt{2}} & 0 \\
0 & 0 & 0 & 0 & 0 & -\tfrac{c}{2\sqrt{2}} & 0 & \tfrac{s}{2\sqrt{2}}
\end{pmatrix}
\label{eq:Ttilde10_app}
\end{equation}

\begin{equation}
\widetilde{\mathcal{T}}^{11} =
\begin{pmatrix}
0 & 0 & 0 & -\tfrac{i}{2\sqrt{2}} & 0 & 0 & 0 & 0 \\
0 & 0 & -\tfrac{i}{2\sqrt{2}} & 0 & 0 & 0 & 0 & 0 \\
0 & \tfrac{i}{2\sqrt{2}} & 0 & 0 & 0 & 0 & 0 & 0 \\
\tfrac{i}{2\sqrt{2}} & 0 & 0 & 0 & 0 & 0 & 0 & 0 \\
0 & 0 & 0 & 0 & 0 & 0 & 0 & -\tfrac{i}{2\sqrt{2}} \\
0 & 0 & 0 & 0 & 0 & 0 & -\tfrac{i}{2\sqrt{2}} & 0 \\
0 & 0 & 0 & 0 & 0 & \tfrac{i}{2\sqrt{2}} & 0 & 0 \\
0 & 0 & 0 & 0 & \tfrac{i}{2\sqrt{2}} & 0 & 0 & 0
\end{pmatrix}
\label{eq:Ttilde11_app}
\end{equation}

\begin{equation}
\widetilde{\mathcal{T}}^{12} =
\begin{pmatrix}
0 & 0 & 0 & -\tfrac{1}{2\sqrt{2}} & 0 & 0 & 0 & 0 \\
0 & 0 & \tfrac{1}{2\sqrt{2}} & 0 & 0 & 0 & 0 & 0 \\
0 & \tfrac{1}{2\sqrt{2}} & 0 & 0 & 0 & 0 & 0 & 0 \\
-\tfrac{1}{2\sqrt{2}} & 0 & 0 & 0 & 0 & 0 & 0 & 0 \\
0 & 0 & 0 & 0 & 0 & 0 & 0 & \tfrac{1}{2\sqrt{2}} \\
0 & 0 & 0 & 0 & 0 & 0 & -\tfrac{1}{2\sqrt{2}} & 0 \\
0 & 0 & 0 & 0 & 0 & -\tfrac{1}{2\sqrt{2}} & 0 & 0 \\
0 & 0 & 0 & 0 & \tfrac{1}{2\sqrt{2}} & 0 & 0 & 0
\end{pmatrix}
\label{eq:Ttilde12_app}
\end{equation}

\begin{equation}
\widetilde{\mathcal{T}}^{13} =
\begin{pmatrix}
0 & 0 & -\tfrac{i}{2\sqrt{2}} & 0 & 0 & 0 & 0 & 0 \\
0 & 0 & 0 & \tfrac{i}{2\sqrt{2}} & 0 & 0 & 0 & 0 \\
\tfrac{i}{2\sqrt{2}} & 0 & 0 & 0 & 0 & 0 & 0 & 0 \\
0 & -\tfrac{i}{2\sqrt{2}} & 0 & 0 & 0 & 0 & 0 & 0 \\
0 & 0 & 0 & 0 & 0 & 0 & -\tfrac{i}{2\sqrt{2}} & 0 \\
0 & 0 & 0 & 0 & 0 & 0 & 0 & \tfrac{i}{2\sqrt{2}} \\
0 & 0 & 0 & 0 & \tfrac{i}{2\sqrt{2}} & 0 & 0 & 0 \\
0 & 0 & 0 & 0 & 0 & -\tfrac{i}{2\sqrt{2}} & 0 & 0
\end{pmatrix}
\label{eq:Ttilde13_app}
\end{equation}

\begin{equation}
\widetilde{\mathcal{T}}^{14} =
\begin{pmatrix}
0 & 0 & -\tfrac{i}{2\sqrt{2}} & 0 & 0 & 0 & 0 & 0 \\
0 & 0 & 0 & -\tfrac{i}{2\sqrt{2}} & 0 & 0 & 0 & 0 \\
\tfrac{i}{2\sqrt{2}} & 0 & 0 & 0 & 0 & 0 & 0 & 0 \\
0 & \tfrac{i}{2\sqrt{2}} & 0 & 0 & 0 & 0 & 0 & 0 \\
0 & 0 & 0 & 0 & 0 & 0 & -\tfrac{i}{2\sqrt{2}} & 0 \\
0 & 0 & 0 & 0 & 0 & 0 & 0 & -\tfrac{i}{2\sqrt{2}} \\
0 & 0 & 0 & 0 & \tfrac{i}{2\sqrt{2}} & 0 & 0 & 0 \\
0 & 0 & 0 & 0 & 0 & \tfrac{i}{2\sqrt{2}} & 0 & 0
\end{pmatrix}
\label{eq:Ttilde14_app}
\end{equation}

\begin{equation}
\widetilde{\mathcal{T}}^{15} =
\begin{pmatrix}
\tfrac{c}{2\sqrt{2}} & 0 & -\tfrac{s}{2\sqrt{2}} & 0 & 0 & 0 & 0 & 0 \\
0 & \tfrac{c}{2\sqrt{2}} & 0 & -\tfrac{s}{2\sqrt{2}} & 0 & 0 & 0 & 0 \\
-\tfrac{s}{2\sqrt{2}} & 0 & -\tfrac{c}{2\sqrt{2}} & 0 & 0 & 0 & 0 & 0 \\
0 & -\tfrac{s}{2\sqrt{2}} & 0 & -\tfrac{c}{2\sqrt{2}} & 0 & 0 & 0 & 0 \\
0 & 0 & 0 & 0 & -\tfrac{c}{2\sqrt{2}} & 0 & \tfrac{s}{2\sqrt{2}} & 0 \\
0 & 0 & 0 & 0 & 0 & -\tfrac{c}{2\sqrt{2}} & 0 & \tfrac{s}{2\sqrt{2}} \\
0 & 0 & 0 & 0 & \tfrac{s}{2\sqrt{2}} & 0 & \tfrac{c}{2\sqrt{2}} & 0 \\
0 & 0 & 0 & 0 & 0 & \tfrac{s}{2\sqrt{2}} & 0 & \tfrac{c}{2\sqrt{2}}
\end{pmatrix}
\label{eq:Ttilde15_app}
\end{equation}

All fifteen generators reduce to their unrotated counterparts $\mathcal{T}_1$--$\mathcal{T}_{15}$
at $\theta = 0$, as required.
The TC limit $\theta = \pi/2$ gives $c = 0$, $s = 1$,
$c_{1/2} = s_{1/2} = 1/\sqrt{2}$.

\subsection*{A.5\quad Linear combinations and commutation relations of the $\SU{4}_V$ generators}

Having the full set of fifteen misaligned generators explicitly, we now organize them into
the physically meaningful linear combinations that diagonalize the action of the
custodial symmetry $\SU{2}_V$.
The starting point is that the fifteen generators $\widetilde{\mathcal{T}}_1$--$\widetilde{\mathcal{T}}_{15}$
do not individually transform as irreducible representations of $\SU{2}_V$; it is
their specific linear combinations that do.
From the first six generators one forms the vector and axial isotriplets,
\begin{equation}
  T^{V,(a)} = \frac{1}{\sqrt{2}}(\widetilde{\mathcal{T}}^a + \widetilde{\mathcal{T}}^{a+3}),
  \qquad
  T^{A,(a)} = \frac{1}{\sqrt{2}}(\widetilde{\mathcal{T}}^a - \widetilde{\mathcal{T}}^{a+3}),
  \qquad a = 1,2,3.
  \label{eq:TVA_def_app}
\end{equation}
The combination $T^{V,(a)}$ is the custodial $\SU{2}_V$ generator satisfying
$[T^{V,(a)}, T^{V,(b)}] = \frac{i}{\sqrt{2}}\eps_{abc}T^{V,(c)}$, and because the $\theta$-dependent
parts of $\widetilde{\mathcal{T}}^a$ and $\widetilde{\mathcal{T}}^{a+3}$ enter with
opposite signs they cancel identically in the sum, making $T^{V,(a)}$
independent of the misalignment angle $\theta$.
The axial combination $T^{A,(a)}$ retains the $\theta$-dependent terms and
its coupling strengths interpolate between the TC and CH limits.
The remaining nine generators $\widetilde{\mathcal{T}}_7$--$\widetilde{\mathcal{T}}_{15}$
are already in irreducible form under $\SU{2}_V$, and are named as
\begin{equation}
\begin{aligned}
  X^{I,(i)}   &= \bigl(\widetilde{\mathcal{T}}^7,\,\widetilde{\mathcal{T}}^8,\,\widetilde{\mathcal{T}}^9\bigr),
  \qquad & S^{I}  &= \widetilde{\mathcal{T}}^{10}, \\[4pt]
  X^{II,(i)}  &= \bigl(\widetilde{\mathcal{T}}^{11},\,\widetilde{\mathcal{T}}^{12},\,\widetilde{\mathcal{T}}^{13}\bigr),
  \qquad & S^{II} &= \widetilde{\mathcal{T}}^{14}, \\[4pt]
  S^{0}     &= \widetilde{\mathcal{T}}^{15}.
\end{aligned}
\label{eq:naming_app}
\end{equation}
Under $\SU{2}_V$ the triplets $X^{I,(i)}$ and $X^{II,(i)}$ each transform as
$[T^{V,(a)}, X^{(\alpha = I,II),i}] = \frac{i}{\sqrt{2}}\eps_{aij}X^{(\alpha),j}$, while the singlets
$S^{I}$, $S^{II}$, and $S^{0}$ commute with all $T^{V,(a)}$.
The complete set of commutation relations between all $\SU{4}_V$ generators is collected in
Table~\ref{tab:commutators_app}, and similarly for the broken generators, the commutation relations are summarized in Table~\ref{tab:broken-generators}.
Inserting these generators into the current formula of
Eq.~\eqref{eq:current_formula} gives the hyperquark currents of the fifteen spin-1
resonances, which are classified in Table~\ref{tab:currents_app} according to the
$\SU{2}_V$ representation of the associated generator; the three isotriplet currents
of its first block are those quoted in Eq.~\eqref{eq:rho_iso_currents} of the main
text.

\medskip

\begin{table}[!htbp]
\small
\centering
\caption{Commutation relations of the $\SU{4}_V$ generators in the basis
of Eqs.~\eqref{eq:TVA_def_app} and \eqref{eq:naming_app}, organized by the $\SU{2}_V$ transformations of the generators.
The indices $a,b,c$ and $i,j,k$ both run over $1,2,3$, with $\eps_{abc}$ the Levi-Civita symbol.}
\label{tab:commutators_app}
\renewcommand{\arraystretch}{1.6}
\thinmuskip=3mu
\medmuskip=4mu
\thickmuskip=5mu
\begin{tabular}{>{\raggedright\arraybackslash}p{4.2cm} c >{\raggedright\arraybackslash}p{8.5cm}}
\hline
\textbf{Generators} & $\mathbf{\SU{2}_V}$ & \textbf{Commutation relations} \\
\hline\hline
$T^{V,(a)} = \tfrac{1}{\sqrt{2}}(\widetilde{\mathcal{T}}^a + \widetilde{\mathcal{T}}^{a+3})$
& $\mathbf{3}$
& $[T^{V,(a)},T^{V,(b)}] = \tfrac{i}{\sqrt{2}}\eps_{abc}T^{V,(c)}$,\quad
  $[T^{V,(a)},T^{A,(b)}] = \tfrac{i}{\sqrt{2}}\eps_{abc}T^{A,(c)}$ \\
\hline
$T^{A,(a)} = \tfrac{1}{\sqrt{2}}(\widetilde{\mathcal{T}}^a - \widetilde{\mathcal{T}}^{a+3})$
& $\mathbf{3}$
& $[T^{A,(a)},T^{A,(b)}] = \tfrac{i}{\sqrt{2}}\eps_{abc}T^{V,(c)}$,\quad
  $[T^{A,(a)},X^{\mathrm{I},(i)}] = \tfrac{i}{\sqrt{2}}\delta_{ai}S^{\mathrm{II}}$,\quad
  $[T^{A,(a)},X^{\mathrm{II},(i)}] = -\tfrac{i}{\sqrt{2}}\delta_{ai}S^{\mathrm{I}}$,\quad
  $[T^{A,(a)},S^{\mathrm{I}}] = \tfrac{i}{\sqrt{2}}X^{\mathrm{II},(a)}$,\quad
  $[T^{A,(a)},S^{\mathrm{II}}] = -\tfrac{i}{\sqrt{2}}X^{\mathrm{I},(a)}$ \\
\hline
$X^{\mathrm{I},(i)} = (\widetilde{\mathcal{T}}^7,\widetilde{\mathcal{T}}^8,\widetilde{\mathcal{T}}^9)$,\quad
$S^{\mathrm{I}} = \widetilde{\mathcal{T}}^{10}$
& $\mathbf{3}\oplus\mathbf{1}$
& $[T^{V,(a)},X^{\mathrm{I},(i)}] = \tfrac{i}{\sqrt{2}}\eps_{aij}X^{\mathrm{I},(j)}$,\quad
  $[T^{V,(a)},S^{\mathrm{I}}] = 0$,\quad
  $[X^{\mathrm{I},(i)},X^{\mathrm{I},(j)}] = \tfrac{i}{\sqrt{2}}\eps_{ijk}T^{V,(k)}$,\quad
  $[X^{\mathrm{I},(i)},X^{\mathrm{II},(j)}] = \tfrac{i}{\sqrt{2}}\delta_{ij}S^{0}$,\quad
  $[S^{\mathrm{II}},X^{\mathrm{I},(i)}] = -\tfrac{i}{\sqrt{2}}T^{A,(i)}$ \\
\hline
$X^{\mathrm{II},(i)} = (\widetilde{\mathcal{T}}^{11},\widetilde{\mathcal{T}}^{12},\widetilde{\mathcal{T}}^{13})$,\quad
$S^{\mathrm{II}} = \widetilde{\mathcal{T}}^{14}$
& $\mathbf{3}\oplus\mathbf{1}$
& $[T^{V,(a)},X^{\mathrm{II},(i)}] = \tfrac{i}{\sqrt{2}}\eps_{aij}X^{\mathrm{II},(j)}$,\quad
  $[T^{V,(a)},S^{\mathrm{II}}] = 0$,\quad
  $[X^{\mathrm{II},(i)},X^{\mathrm{II},(j)}] = \tfrac{i}{\sqrt{2}}\eps_{ijk}T^{V,(k)}$,\quad
  $[S^{\mathrm{I}},X^{\mathrm{II},(i)}] = \tfrac{i}{\sqrt{2}}T^{A,(i)}$ \\
\hline
$S^{0} = \widetilde{\mathcal{T}}^{15}$
& $\mathbf{1}$
& $[T^{V,(a)},S^{0}] = 0$,\quad
  $[S^{\mathrm{I}},S^{\mathrm{II}}] = \tfrac{i}{\sqrt{2}}S^{0}$,\quad
  $[S^{\mathrm{I}},S^{0}] = -\tfrac{i}{\sqrt{2}}S^{\mathrm{II}}$,\quad
  $[S^{\mathrm{II}},S^{0}] = \tfrac{i}{\sqrt{2}}S^{\mathrm{I}}$,\quad
  $[S^{0},X^{\mathrm{I},(i)}] = \tfrac{i}{\sqrt{2}}X^{\mathrm{II},(i)}$,\quad
  $[S^{0},X^{\mathrm{II},(i)}] = -\tfrac{i}{\sqrt{2}}X^{\mathrm{I},(i)}$ \\
\hline\hline
\multicolumn{3}{l}{$T^{V,(a)}$, $X^{\mathrm{II},(i)}$, and $S^{\mathrm{II}}$ are independent of
  the vacuum misalignment angle $\theta$.} \\
\hline
\end{tabular}
\end{table}

\begin{table}[!htbp]
\small
\centering
\caption{Decomposition of the broken generators $\widetilde{\mathcal{X}}^a$ under custodial symmetry $[\SU{2}_V]_{(\SU{2}_L,\SU{2}_R)}$, with their commutation relations onto the unbroken generators $\widetilde{\mathcal{T}}^a$. The triplet $K^{V,(a)}$ is associated with the light pions $\pi^{V,(a)}$, identified as the EW bosons. Superscript $S$ marks the $\SU{2}_V$ singlet members $K^{S,\mathrm{I}}$, $K^{S,\mathrm{II}}$, $K^{S,0}$; the remaining $K$'s are triplets. The indices $a,b,c$ and $i,j,k$ both run over $1,2,3$, with $\eps_{abc}$ the Levi-Civita symbol.}
\label{tab:broken-generators}
\renewcommand{\arraystretch}{1.6}
\begin{tabular}{>{\raggedright\arraybackslash}p{4.2cm} c >{\raggedright\arraybackslash}p{8.5cm}}
\hline
\textbf{Generators} & $\mathbf{\SU{2}_V}$ & \textbf{Commutation relations} \\
\hline\hline
$K^{V,(a)} = \tfrac{1}{\sqrt{2}}(\widetilde{\mathcal{X}}^a + \widetilde{\mathcal{X}}^{a+3})$
& $\mathbf{3}$
& $[K^{V,(a)},K^{V,(b)}] = \tfrac{i}{\sqrt{2}}\eps_{abc}T^{V,(c)}$,\quad
  $[K^{V,(a)},K^{A,(b)}] = \tfrac{i}{\sqrt{2}}\eps_{abc}T^{A,(c)}$,\quad
  $[K^{V,(a)},K^{\mathrm{I},(i)}] = \tfrac{i}{\sqrt{2}}\eps_{aic}X^{\mathrm{I},(c)}$,\quad
  $[K^{V,(a)},K^{\mathrm{II},(i)}] = \tfrac{i}{\sqrt{2}}\eps_{aic}X^{\mathrm{II},(c)}$ \\
\hline
$K^{A,(a)} = \tfrac{1}{\sqrt{2}}(\widetilde{\mathcal{X}}^a - \widetilde{\mathcal{X}}^{a+3})$
& $\mathbf{3}$
& $[K^{A,(a)},K^{A,(b)}] = \tfrac{i}{\sqrt{2}}\eps_{abc}T^{V,(c)}$,\quad
  $[K^{A,(a)},K^{\mathrm{I},(i)}] = \tfrac{i}{\sqrt{2}}\delta_{ai}S^{\mathrm{II}}$,\quad
  $[K^{A,(a)},K^{\mathrm{II},(i)}] = -\tfrac{i}{\sqrt{2}}\delta_{ai}S^{\mathrm{I}}$,\quad
  $[K^{A,(a)},K^{S,\mathrm{I}}] = \tfrac{i}{\sqrt{2}}X^{\mathrm{II},(a)}$,\quad
  $[K^{A,(a)},K^{S,\mathrm{II}}] = -\tfrac{i}{\sqrt{2}}X^{\mathrm{I},(a)}$ \\
\hline
$K^{\mathrm{I},(i)} = (\widetilde{\mathcal{X}}^7,\widetilde{\mathcal{X}}^8,\widetilde{\mathcal{X}}^9)$,\quad
$K^{S,\mathrm{I}} = \widetilde{\mathcal{X}}^{10}$
& $\mathbf{3}\oplus\mathbf{1}$
& $[T^{V,(a)},K^{\mathrm{I},(i)}] = \tfrac{i}{\sqrt{2}}\eps_{aij}K^{\mathrm{I},(j)}$,\quad
  $[T^{V,(a)},K^{S,\mathrm{I}}] = 0$,\quad
  $[K^{\mathrm{I},(i)},K^{\mathrm{I},(j)}] = \tfrac{i}{\sqrt{2}}\eps_{ijk}T^{V,(k)}$,\quad
  $[K^{\mathrm{I},(i)},K^{\mathrm{II},(j)}] = \tfrac{i}{\sqrt{2}}\delta_{ij}S^{0}$,\quad
  $[K^{\mathrm{I},(i)},K^{S,\mathrm{II}}] = \tfrac{i}{\sqrt{2}}T^{A,(i)}$ \\
\hline
$K^{\mathrm{II},(i)} = (\widetilde{\mathcal{X}}^{11},\widetilde{\mathcal{X}}^{12},\widetilde{\mathcal{X}}^{13})$,\quad
$K^{S,\mathrm{II}} = \widetilde{\mathcal{X}}^{14}$
& $\mathbf{3}\oplus\mathbf{1}$
& $[T^{V,(a)},K^{\mathrm{II},(i)}] = \tfrac{i}{\sqrt{2}}\eps_{aij}K^{\mathrm{II},(j)}$,\quad
  $[T^{V,(a)},K^{S,\mathrm{II}}] = 0$,\quad
  $[K^{\mathrm{II},(i)},K^{\mathrm{II},(j)}] = \tfrac{i}{\sqrt{2}}\eps_{ijk}T^{V,(k)}$,\quad
  $[K^{S,\mathrm{I}},K^{\mathrm{II},(i)}] = \tfrac{i}{\sqrt{2}}T^{A,(i)}$ \\
\hline
$K^{S,0} = \widetilde{\mathcal{X}}^{15}$
& $\mathbf{1}$
& $[T^{V,(a)},K^{S,0}] = 0$,\quad
  $[K^{S,\mathrm{I}},K^{S,\mathrm{II}}] = \tfrac{i}{\sqrt{2}}S^{0}$,\quad
  $[K^{S,0},K^{S,\mathrm{I}}] = \tfrac{i}{\sqrt{2}}S^{\mathrm{II}}$,\quad
  $[K^{S,0},K^{S,\mathrm{II}}] = -\tfrac{i}{\sqrt{2}}S^{\mathrm{I}}$,\quad
  $[K^{S,0},K^{\mathrm{I},(i)}] = \tfrac{i}{\sqrt{2}}X^{\mathrm{II},(i)}$,\quad
  $[K^{S,0},K^{\mathrm{II},(i)}] = -\tfrac{i}{\sqrt{2}}X^{\mathrm{I},(i)}$ \\
\hline
\end{tabular}
\end{table}

\begin{table}[!htbp]
\small
\centering
\caption{%
  Classification of the fifteen spin-1 resonances in terms of the underlying  hyperquark currents. The states are associated with the unbroken generators of $\SU{4}_V$ in the $\SU{2}_V$ basis of
  Eqs.~\eqref{eq:TVA_def_app} and~\eqref{eq:naming_app}. Here
  $c\equiv\cos\theta$ and $s\equiv\sin\theta$; the states without a hat are $\theta$-independent.
  The three isotriplet currents of the first block are given explicitly in
  Eq.~\eqref{eq:rho_iso_currents} of the main text.}

\label{tab:currents_app}
\renewcommand{\arraystretch}{1.50}
\setlength{\tabcolsep}{6pt}
\begin{tabular}{!{\vrule width 1.5pt} l !{\vrule width 1pt} l !{\vrule width 1pt} c !{\vrule width 1pt} l !{\vrule width 1.5pt}}
\noalign{\hrule height 1.5pt}
\quad\textbf{Generator}
  & \quad\textbf{State}
  & $\boldsymbol{\mathrm{SU}(2)_V}$
  & \quad\textbf{Current}~$\boldsymbol{\rho^\mu}$
\\
\noalign{\hrule height 1.5pt}
\multirow{3}{*}{\quad$T^{V,(a)}$}
  & \quad$\rho_{\rm HC}^{+}$ & \multirow{3}{*}{$\mathbf{3}$}
  & $\dfrac{1}{2}\!\left(\bar{U}_1\gamma^\mu D_1 + \bar{U}_2\gamma^\mu D_2\right)$
\\[8pt]
  & \quad$\rho_{\rm HC}^{-}$ &
  & $\dfrac{1}{2}\!\left(\bar{D}_1\gamma^\mu U_1 + \bar{D}_2\gamma^\mu U_2\right)$
\\[8pt]
  & \quad$\rho_{\rm HC}^{0}$ &
  & $\dfrac{1}{2\sqrt{2}}\!\left(\bar{U}_1\gamma^\mu U_1 + \bar{U}_2\gamma^\mu U_2
      - \bar{D}_1\gamma^\mu D_1 - \bar{D}_2\gamma^\mu D_2\right)$
\\[6pt]
\noalign{\hrule height 1pt}
\multirow{5}{*}{\quad$T^{A,(a)}$}
  & \quad$\hat{\rho}_{\rm HC}^{A,+}$ & \multirow{3}{*}{$\mathbf{3}$}
  & $-\dfrac{1}{2}\!\Big[
      \left(-\bar{U}_1\gamma^\mu D_1+\bar{U}_2\gamma^\mu D_2\right)c
      +\left(\bar{U}_1\gamma^\mu D_2+\bar{U}_2\gamma^\mu D_1\right)s\Big]$
\\[8pt]
  & \quad$\hat{\rho}_{\rm HC}^{A,-}$ &
  & $-\dfrac{1}{2}\!\Big[
      \left(-\bar{D}_1\gamma^\mu U_1+\bar{D}_2\gamma^\mu U_2\right)c
      +\left(\bar{D}_2\gamma^\mu U_1+\bar{D}_1\gamma^\mu U_2\right)s\Big]$
\\[8pt]
  & \quad$\hat{\rho}_{\rm HC}^{A,0}$ &
  & $\dfrac{1}{2\sqrt{2}}\!\Big[\left(-\bar{D}_1\gamma^\mu D_1+\bar{D}_2\gamma^\mu D_2
       +\bar{U}_1\gamma^\mu U_1-\bar{U}_2\gamma^\mu U_2\right)c$
\\[-2pt]
  & & &
  \quad$+\left(\bar{D}_2\gamma^\mu D_1+\bar{D}_1\gamma^\mu D_2
       -\bar{U}_2\gamma^\mu U_1-\bar{U}_1\gamma^\mu U_2\right)s\Big]$
\\[6pt]
\noalign{\hrule height 1pt}
\multirow{7}{*}{\quad$X^{\mathrm{I},(i)},\,S^{\mathrm{I}}$}
  & \quad$\hat{\rho}_{\rm HC}^{X^{\mathrm{I}},+}$ & \multirow{3}{*}{$\mathbf{3}$}
  & $\dfrac{1}{2}\!\Big[
      \left(\bar{U}_1\gamma^\mu D_2+\bar{U}_2\gamma^\mu D_1\right)c
      +\left(\bar{U}_1\gamma^\mu D_1-\bar{U}_2\gamma^\mu D_2\right)s\Big]$
\\[8pt]
  & \quad$\hat{\rho}_{\rm HC}^{X^{\mathrm{I}},-}$ &
  & $\dfrac{1}{2}\!\Big[
      \left(\bar{D}_2\gamma^\mu U_1+\bar{D}_1\gamma^\mu U_2\right)c
      +\left(\bar{D}_1\gamma^\mu U_1-\bar{D}_2\gamma^\mu U_2\right)s\Big]$
\\[8pt]
  & \quad$\hat{\rho}_{\rm HC}^{X^{\mathrm{I}},0}$ &
  & $\dfrac{1}{2\sqrt{2}}\!\Big[\left(-\bar{D}_1\gamma^\mu D_2-\bar{D}_2\gamma^\mu D_1
      +\bar{U}_1\gamma^\mu U_2+\bar{U}_2\gamma^\mu U_1\right)c$
\\[-2pt]
  & & &
  \quad$+\left(-\bar{D}_1\gamma^\mu D_1+\bar{D}_2\gamma^\mu D_2
      +\bar{U}_1\gamma^\mu U_1-\bar{U}_2\gamma^\mu U_2\right)s\Big]$
\\[8pt]
\cline{2-4}
  & \quad$\hat{\mathcal{X}}_{\rm HC}^{S^{\mathrm{I}}}$ & $\mathbf{1}$
  & $\dfrac{1}{2\sqrt{2}}\!\Big[\left(\bar{D}_1\gamma^\mu D_2+\bar{D}_2\gamma^\mu D_1
      +\bar{U}_1\gamma^\mu U_2+\bar{U}_2\gamma^\mu U_1\right)c$
\\[-2pt]
  & & &
  \quad$+\left(\bar{D}_1\gamma^\mu D_1-\bar{D}_2\gamma^\mu D_2
      +\bar{U}_1\gamma^\mu U_1-\bar{U}_2\gamma^\mu U_2\right)s\Big]$
\\[6pt]
\noalign{\hrule height 1pt}
\multirow{5}{*}{\quad$X^{\mathrm{II},(i)},\,S^{\mathrm{II}}$}
  & \quad$\rho_{\rm HC}^{X^{\mathrm{II}},+}$ & \multirow{3}{*}{$\mathbf{3}$}
  & $\dfrac{i}{2}\!\left(-\bar{U}_1\gamma^\mu D_2+\bar{U}_2\gamma^\mu D_1\right)$
\\[8pt]
  & \quad$\rho_{\rm HC}^{X^{\mathrm{II}},-}$ &
  & $\dfrac{i}{2}\!\left(\bar{D}_2\gamma^\mu U_1-\bar{D}_1\gamma^\mu U_2\right)$
\\[8pt]
  & \quad$\rho_{\rm HC}^{X^{\mathrm{II}},0}$ &
  & $\dfrac{i}{2\sqrt{2}}\!\left(\bar{D}_1\gamma^\mu D_2-\bar{D}_2\gamma^\mu D_1
      -\bar{U}_1\gamma^\mu U_2+\bar{U}_2\gamma^\mu U_1\right)$
\\[8pt]
\cline{2-4}
  & \quad$\rho_{\rm HC}^{S^{\mathrm{II}}}$ & $\mathbf{1}$
  & $\dfrac{i}{2\sqrt{2}}\!\left(-\bar{D}_1\gamma^\mu D_2+\bar{D}_2\gamma^\mu D_1
      -\bar{U}_1\gamma^\mu U_2+\bar{U}_2\gamma^\mu U_1\right)$
\\[6pt]
\noalign{\hrule height 1pt}
\quad$S^{0}$
  & \quad$\hat{\mathcal{X}}_{\rm HC}^{S^{0}}$ & $\mathbf{1}$
  & $\dfrac{1}{2\sqrt{2}}\!\Big[\left(\bar{D}_1\gamma^\mu D_1-\bar{D}_2\gamma^\mu D_2
      +\bar{U}_1\gamma^\mu U_1-\bar{U}_2\gamma^\mu U_2\right)c$
\\[-2pt]
  & & &
  \quad$-\left(\bar{D}_2\gamma^\mu D_1+\bar{D}_1\gamma^\mu D_2
      +\bar{U}_2\gamma^\mu U_1+\bar{U}_1\gamma^\mu U_2\right)s\Big]$
\\[6pt]
\noalign{\hrule height 1.5pt}
\end{tabular}
\end{table}
\end{appendix}

\section{HLS Interaction Terms Derivation}
\label{app:HLS}

\subsection*{B.1\quad Chiral fields, one-forms, and the $\rhoT\PiT\PiT$ vertex}
The techni-pions are described by the non-linearly realized field
\begin{equation}
  U \;=\; \exp\!\left(\frac{2i\Pi}{\FPi}\right),
  \qquad
  \Pi = \PiT^a\,\widetilde{\Lambda}^a \qquad a = 1,\dots,15, 
  \label{eq:U_def_app}
\end{equation}
transforming as $U \to g_L\,U\,g_R^\dagger$ under the global chiral symmetry,
with leading-order Lagrangian
\begin{equation}
  \mathcal{L}_U
  \;=\;
  \frac{\FPi^2}{4}\,\tr\bigl[\partial^\mu U^\dagger\,\partial_\mu U\bigr]
  \;=\;
  \half\,\partial^\mu\PiT^a\,\partial_\mu\PiT^a
  \;+\;\dots,
  \label{eq:LU_app}
\end{equation}
where the second equality uses the $4\times4$ generator normalization
$\tr[\widetilde{\Lambda}^a\widetilde{\Lambda}^b] = \tfrac{1}{2}\delta^{ab}$ of
Appendix~\ref{app:generators}, adopted throughout this appendix; all traces here are
taken over the $4\times4$ flavour space.
In the HLS construction $U$ is split into the chiral fields
\begin{equation}
  U = \xi_L^\dagger\,\xi_R,
  \qquad
  \xi_L = e^{-i\Pi/\FPi},
  \quad
  \xi_R = e^{+i\Pi/\FPi},
  \label{eq:xi_def_app}
\end{equation}
where we have fixed the unitary gauge of the hidden local symmetry, in which
the additional scalar degrees of freedom introduced by the decomposition are
set to zero~\cite{Bando:1987br,Harada:2003jx}.
From $\xi_{L,R}$, we define the covariantized Maurer--Cartan one-forms
\begin{equation}
\omega^{\mu}_{L,R} = \frac{1}{2i}\,(D^{\mu}\xi_{L,R})\,\xi^{\dagger}_{L,R},
\end{equation}
where $D^\mu\xi_{L,R} = \partial^\mu\xi_{L,R} - i\gr\,\rhoT^\mu\,\xi_{L,R}$ and
$\rhoT^\mu = \rhoT^{a\mu}\widetilde{\Lambda}^a$ (with $a = 1,\dots,15$) is the
gauge field of $\mathcal{H}_1$. Projecting $\omega^{\mu}_{L,R}$ onto the unbroken
($T_L = T_R$) and broken ($T_L = -T_R$) directions gives
\begin{equation}
  V^\mu = 2\,\mathrm{tr}\bigl[\widetilde{\Lambda}^{a}
    (\omega^{\mu}_{L} + \omega^{\mu}_{R})\bigr]\,\widetilde{\Lambda}^{a},
    \qquad
    A^\mu = 2\,\mathrm{tr}\bigl[\widetilde{\Lambda}^{a}
    (\omega^{\mu}_{L} - \omega^{\mu}_{R})\bigr]\,\widetilde{\Lambda}^{a},
\end{equation}
which reduce to
\begin{equation}
  V^\mu = \frac{1}{2i}\bigl[(D^\mu\xi_L)\,\xi_L^\dagger
           + (D^\mu\xi_R)\,\xi_R^\dagger\bigr],
  \qquad
  A^\mu = \frac{1}{2i}\bigl[(D^\mu\xi_L)\,\xi_L^\dagger
           - (D^\mu\xi_R)\,\xi_R^\dagger\bigr].
  \label{eq:oneforms_app}
\end{equation}
The factor of two is the inverse of
$\tr[\widetilde{\Lambda}^a\widetilde{\Lambda}^b]=\tfrac12\delta^{ab}$, and ensures
that the projection returns the one-forms themselves.

We first establish the transformation properties of these objects. Under the
transformations, $\xi_{L,R}\to h(x)\,\xi_{L,R}\,g_{L,R}^\dagger$
and
$\rhoT^\mu \to h\,\rhoT^\mu\,h^\dagger + \tfrac{i}{\gr}\,h\,\partial^\mu h^\dagger$,
the covariant derivative transforms as
\begin{align}
  D^\mu\xi_{L,R}
  \;\to\;&
  \partial^\mu\!\bigl(h\,\xi_{L,R}\,g_{L,R}^\dagger\bigr)
  - i\gr\left[h\,\rhoT^\mu\,h^\dagger
  + \frac{i}{\gr}\,h\,\partial^\mu h^\dagger\right]
  h\,\xi_{L,R}\,g_{L,R}^\dagger
  \nonumber\\[4pt]
  =\;&
  (\partial^\mu h)\,\xi_{L,R}\,g_{L,R}^\dagger
  + h\,(\partial^\mu\xi_{L,R})\,g_{L,R}^\dagger
  - i\gr\,h\,\rhoT^\mu\,\xi_{L,R}\,g_{L,R}^\dagger
  + h\,(\partial^\mu h^\dagger)\,h\,\xi_{L,R}\,g_{L,R}^\dagger
  \nonumber\\[4pt]
  =\;&
  h\,\bigl(D^\mu\xi_{L,R}\bigr)\,g_{L,R}^\dagger,
  \label{eq:covder_transf_app}
\end{align}
where we used $h^\dagger h = 1$ and
$h\,(\partial^\mu h^\dagger)\,h = -\,\partial^\mu h$, which follows from
differentiating $h\,h^\dagger = 1$. The inhomogeneous term of the $\rhoT$
transformation thus cancels the derivative of the local factor $h(x)$.
Multiplying from the right by $\xi_{L,R}^\dagger \to g_{L,R}\,\xi_{L,R}^\dagger\,h^\dagger$,
the global factor cancels, $g_{L,R}^\dagger\,g_{L,R} = 1$, giving
\begin{equation}
  (D^\mu\xi_{L,R})\,\xi_{L,R}^\dagger
  \;\to\;
  h\,\bigl[(D^\mu\xi_{L,R})\,\xi_{L,R}^\dagger\bigr]\,h^\dagger,
  \label{eq:oneform_transf_app}
\end{equation}
so that both one-forms transform homogeneously,
$V^\mu \to h\,V^\mu h^\dagger$ and $A^\mu \to h\,A^\mu h^\dagger$, and the
traces of their squares in the HLS Lagrangian~\eqref{eq:HLS} are invariant.

We now expand the one-forms in powers of $\Pi$. Expanding
Eq.~\eqref{eq:xi_def_app} to second order,
\begin{align}
  \xi_L \;=\; 1 - \frac{i\Pi}{\FPi} - \frac{\Pi^2}{2\FPi^2}
               + \mathcal{O}(\Pi^3),
  \qquad
  \xi_L^\dagger \;=\; 1 + \frac{i\Pi}{\FPi} - \frac{\Pi^2}{2\FPi^2}
                  + \mathcal{O}(\Pi^3),
\end{align}
and similarly $\xi_R = 1 + i\Pi/\FPi - \Pi^2/(2\FPi^2)+\cdots$,
$\xi_R^\dagger = 1 - i\Pi/\FPi - \Pi^2/(2\FPi^2)+\cdots$.
The covariant derivative of $\xi_L$ reads
\begin{equation}
  D^\mu\xi_L
  = -\frac{i\,\partial^\mu\Pi}{\FPi}
  - \frac{\partial^\mu\Pi\,\Pi + \Pi\,\partial^\mu\Pi}{2\FPi^2}
  - i\gr\,\rhoT^\mu\,\xi_L
  + \mathcal{O}(\Pi^3).
\end{equation}
Multiplying from the right by $\xi_L^\dagger$, using
$\rhoT^\mu\,\xi_L\,\xi_L^\dagger = \rhoT^\mu$, and retaining terms to order
$\Pi^2$,
\begin{equation}
  (D^\mu\xi_L)\,\xi_L^\dagger
  = -\frac{i\,\partial^\mu\Pi}{\FPi}
  + \frac{[\partial^\mu\Pi,\,\Pi]}{2\FPi^2}
  - i\gr\,\rhoT^{a\mu}\widetilde{\Lambda}^a
  + \mathcal{O}(\Pi^3).
\end{equation}
Performing the same steps for $\xi_R$ gives
\begin{equation}
  (D^\mu\xi_R)\,\xi_R^\dagger
  = +\frac{i\,\partial^\mu\Pi}{\FPi}
  + \frac{[\partial^\mu\Pi,\,\Pi]}{2\FPi^2}
  - i\gr\,\rhoT^{a\mu}\widetilde{\Lambda}^a
  + \mathcal{O}(\Pi^3).
\end{equation}
Using $\Pi = \PiT^b\widetilde{\Lambda}^b$ and the $\SU{4}$ Lie algebra
$[\widetilde{\Lambda}^b,\widetilde{\Lambda}^c] = if^{bca}\,\widetilde{\Lambda}^a$,
the commutator expands as
\begin{equation}
  [\partial^\mu\Pi,\,\Pi]
  = \partial^\mu\PiT^b\,\PiT^c\,
    [\widetilde{\Lambda}^b,\widetilde{\Lambda}^c]
  = i\,f^{bca}\,\partial^\mu\PiT^b\,\PiT^c\,\widetilde{\Lambda}^a.
  \label{eq:comm_app}
\end{equation}
This is precisely where the $\SU{4}$ structure constants $f^{abc}$ enter:
the commutator of two broken directions, $(\widetilde{\Lambda}^b,-\widetilde{\Lambda}^b)$
and $(\widetilde{\Lambda}^c,-\widetilde{\Lambda}^c)$ in the $(T_L,T_R)$ notation of
Appendix~\ref{app:generators}, produces an \emph{unbroken} one,
$if^{bca}(\widetilde{\Lambda}^a,+\widetilde{\Lambda}^a)$, as required by the symmetric coset structure of
$(\SU{4}_L\times\SU{4}_R)/\SU{4}_V$~\cite{Panico:2015jxa}, and the relations are
given in Table~\ref{tab:broken-generators}.
Adding and subtracting according to Eq.~\eqref{eq:oneforms_app} and using
the commutator relation~\eqref{eq:comm_app}, the one-form expansions are
\begin{align}
  V^\mu &= -\,\gr\,\rhoT^{a\mu}\,\widetilde{\Lambda}^a
           \;+\;
           \frac{f^{bca}}{2\FPi^2}\,
           \partial^\mu\PiT^b\,\PiT^c\,\widetilde{\Lambda}^a
           \;+\; \mathcal{O}(\Pi^3),
  \label{eq:Vmu_def} \\[4pt]
  A^\mu &= -\,\frac{1}{\FPi}\,\partial^\mu\PiT^a\,\widetilde{\Lambda}^a
           \;+\; \mathcal{O}(\Pi^2),
  \label{eq:Amu_def}
\end{align}
confirming that $V^\mu$ is valued along the unbroken and $A^\mu$ along the
broken generators.
Substituting the expansion~\eqref{eq:Vmu_def} into $\tr[V^\mu V_\mu]$ and
projecting onto the generator basis,
\begin{align}
   aF^2_{\Pi}\,\tr\!\bigl[V^\mu V_\mu\bigr]
  &= aF^2_{\Pi}\, V^{a\mu}V^b_\mu\,\tr[\widetilde{\Lambda}^a\widetilde{\Lambda}^b]
   = \frac{aF^2_{\Pi}}{2}\,V^{a\mu}V^a_\mu
  \nonumber\\[6pt]
  &= \frac{aF^2_{\Pi}}{2}\gr^2\,\rhoT^{a\mu}\rhoT^a{}_\mu
  \;-\;
    a\gr\,\rhoT^{a\mu}\,
       \frac{f^{bca}}{2}\,
       \partial_\mu\PiT^b\,\PiT^c
  \;+\;\mathcal{O}(\Pi^4).
  \label{eq:VV_exp}
\end{align}
The first term yields the $\rhoT$ mass
$\MrhoT^2 = a\,\gr^2\FPi^2$~\cite{PhysRevLett.16.255,PhysRev.147.1071}, while the cross term, after relabelling
$b\leftrightarrow c$ and using the antisymmetry of $f^{abc}$, gives the
interaction Lagrangian
\begin{equation}
  \mathcal{L}_{\rhoT\PiT\PiT}
  = g_{\rhoT\PiT\PiT}\,f^{abc}\,\rhoT^{a\mu}\,\PiT^b\,\partial_\mu\PiT^c,
  \qquad
  g_{\rhoT\PiT\PiT} = \frac{a\,\gr}{2},
  \label{eq:Lrhopipi}
\end{equation}
which is Eq.~\eqref{eq:rhopipi} of the main text.
Similarly, substituting the expansion~\eqref{eq:Amu_def} into the first term
of the HLS Lagrangian~\eqref{eq:HLS} gives
\begin{equation}
  \FPi^2\,\tr\bigl[A^\mu A_\mu\bigr]
  \;=\;
  \half\,\partial^\mu\PiT^a\,\partial_\mu\PiT^a
  \;+\;\dots,
  \label{eq:AA_kinetic_app}
\end{equation}
reproducing the canonical techni-pion kinetic term of Eq.~\eqref{eq:LU_app}.

\subsection*{B.2\quad Partial decay width}
From the Feynman rule $ig_{\rhoT\PiT\PiT} f^{abc}(p_b - p_c)^\mu$ derived from
Eq.~\eqref{eq:Lrhopipi}, the $P$-wave partial decay width for
$\rhoT^a \to \PiT^b\PiT^c$ follows the standard HLS construction of vector
meson partial widths (QCD: $\rho \to \pi\pi$)~\cite{Bando:1987br,Harada:2003jx},
with the coupling $g_{\rhoT\PiT\PiT}$ generalizing the QCD $\rho\pi\pi$
coupling $g_{\rho\pi\pi}$ and the $\SU{4}$ structure constants $f^{abc}$
generalizing the $\SU{2}$ isospin factor $\varepsilon^{abc}$ of the QCD
vertex,
\begin{equation}
  \Gamma(\rhoT^a \to \PiT^b\PiT^c)
  = \frac{g_{\rhoT\PiT\PiT}^2\,|f^{abc}|^2}{6\pi}\cdot\frac{p^3_{bc}}{\MrhoT^2},
  \label{eq:Gamma_app}
\end{equation}
where $p_{bc}$ is the magnitude of the three-momentum of each final-state
particle in the $\rhoT$ rest frame. For final-state masses $M_b$ and $M_c$ it
is given by
\begin{equation}
  p_{bc}
  \;=\;
  \frac{\MrhoT}{2}\,
  \sqrt{\left(1 - \frac{(M_b + M_c)^2}{\MrhoT^2}\right)
        \left(1 - \frac{(M_b - M_c)^2}{\MrhoT^2}\right)},
  \label{eq:pBC_app}
\end{equation}
which vanishes at the kinematic threshold $\MrhoT = M_b + M_c$. The physical
EW boson masses entering the light-channel phase space are
$M_W \approx 80.4\,\mathrm{GeV}$ and $M_Z \approx 91.2\,\mathrm{GeV}$. For
$\MrhoT = 2\,\mathrm{TeV}$ both factors in Eq.~\eqref{eq:pBC_app} differ from
unity only at the level of $10^{-3}$, so the light channel is taken in the
equal-mass limit with $M_b = M_c = M_W$; the induced error on $\beta^3$ is
below $0.2\%$. The heavy pions are likewise taken to be degenerate with common
mass $M_{\PiT^H}$. Within each channel the momentum therefore reduces to the
equal-mass form
\begin{equation}
  p_{bc} \;=\; \frac{\MrhoT}{2}\,\beta,
  \qquad
  \beta = \sqrt{1 - \frac{4M^2}{\MrhoT^2}},
  \label{eq:beta_app}
\end{equation}
with $M = M_W$ for the light channel and $M = M_{\PiT^H}$ for the heavy
channel. With this choice the $W^+W^-$ and $W^\pm Z$ phase-space factors agree to better
than $0.2\%$ ($0.03\%$) at $\MrhoT = 2\,(5)\,\mathrm{TeV}$, so the two
branching ratios differ by less than $0.04$ percentage points and are shown as
a single curve. Here $\beta$ is the
velocity of the final-state particles and $\beta^3$ the $P$-wave phase-space
factor used in the branching-ratio formulas of Appendix~\ref{app:phasespace}.
\subsection*{B.3\quad The $\omT$ interaction and the WZW anomaly}
The $\omT$ is the gauge boson of the hidden $[\UN{1}_V]_{\rm local}$
symmetry. Its generator $T^0 \propto \mathbb{1}_{4\times4}$ commutes with
all $\SU{4}$ generators, so the structure constants satisfy $f^{bc0} = 0$
identically. Substituting into the HLS vertex~\eqref{eq:rhopipi}, the
$\omT\PiT\PiT$ coupling vanishes at every order in the HLS expansion, and
the $\omT$ cannot decay into two techni-pions through the HLS Lagrangian.
In QCD the isosinglet $\omega$ meson faces the same selection rule and
decays instead through the three-body channel $\omega \to \pi^+\pi^-\pi^0$,
generated by the WZW anomaly
term~\cite{Wess:1971yu,Witten:1983tw,Harada:2003jx,PhysRevD.30.594}. By the same mechanism,
the $\omT$ couples to three techni-pions through the WZW anomaly of the
hypercolor gauge theory,
\begin{equation}
  \mathcal{L}_{\rm WZW}
  \;\supset\;
  \frac{\NHC}{12\pi^2\FPi^3}\,
  \eps_{\mu\nu\rho\sigma}\,\omT^\mu\,
  f^{abc}\,
  \partial^\nu\PiT^a\,\partial^\rho\PiT^b\,
  \partial^\sigma\PiT^c,
  \label{eq:WZW_om_app}
\end{equation}
where $f^{abc}$ are the antisymmetric $\SU{4}$ structure constants. The
totally antisymmetric contraction with $\eps_{\mu\nu\rho\sigma}$ selects the
antisymmetric part of the flavour trace, so the symmetric $d^{abc}$
combinations drop out and each antisymmetric pion triple
$(\PiT^a,\PiT^b,\PiT^c)$ couples with weight $f^{abc}$. Because the broken
and unbroken generators of the $(\SU{4}_L\times\SU{4}_R)/\SU{4}_V$ coset carry
the same $\SU{4}$ adjoint label, the structure constant $f^{abc}$ coupling a
given pion triple is read off directly from the commutation relations of
Tables~\ref{tab:broken-generators} and \ref{tab:commutators_app}, as the
coefficient of $T^a$ in $[K^b,K^c]$, and each non-vanishing entry corresponds
to one three-body channel of the $\omT$, with phase space deciding which
channels are open. Among the light techni-pions, identified with the
longitudinal $W^+$, $Z^0$, $W^-$, the relevant structure constants reduce to
$\eps^{abc}$ of the embedded $\SU{2}_V$, giving the three-body decay
$\omT \to W^+Z^0W^-$ in direct correspondence with the QCD
$\omega \to \pi^+\pi^-\pi^0$~\cite{Chivukula:1989rn}. Beyond this all-light
channel, the same vertex generates decays into mixed triples
$\pi_{\rm HC}^V\PiT^H\PiT^H$ and all-heavy triples $\PiT^H\PiT^H\PiT^H$ for
every non-vanishing $f^{abc}$. This decay pattern is governed entirely by the
topological structure of the hypercolor theory.

\section{Group Weights and Branching Ratios}
\label{app:phasespace}
\subsection*{C.1\quad The $\SU{4}$ group weights and sum rule}
\label{app:groupfactors}
We classify the fifteen techni-pions into two categories. The three light
pions, denoted $\pi_{\rm HC}^V$, are the isotriplet states associated with the
generators $T^{V,(a)}$ with $a=1,2,3$. In the broken phase they are identified
with the longitudinal components of the EW bosons $W^\pm$ and $Z^0$.
The remaining twelve heavy pions, denoted $\PiT^H$, sit near the EW scale. The
group-theoretic weight for the decay of $\rhoT^a$ into each category is
encoded in the group weights
\begin{equation}
  C_L^a \;\equiv\; \sum_{b,c\,\in\,\pi_{\rm HC}^V} |f^{abc}|^2,
  \qquad
  C_H^a \;\equiv\; \sum_{b,c\,\in\,\PiT^H} |f^{abc}|^2,
  \qquad
  C_\mathrm{mix}^a \;\equiv\; \sum_{\substack{b\,\in\,\pi_{\rm HC}^V \\ c\,\in\,\PiT^H}}
  2\,|f^{abc}|^2,
  \label{eq:group_factors_app}
\end{equation}
where the sums run over ordered index pairs, the factor of $2$ in
$C_\mathrm{mix}^a$ accounting for both orderings of the light and heavy
indices.

For $\SU{N}$ the structure constants satisfy a sum rule that follows from the
quadratic Casimir of the adjoint representation~\cite{Georgi:1999wka}. Defining
the adjoint representation matrices by
$(T^b_\mathrm{adj})_{ac} = -if^{bac}$, the Casimir relation
$\sum_b T^b_\mathrm{adj}T^b_\mathrm{adj} = C_2(\mathrm{adj})\,\mathbb{1}$
becomes, in components,
\begin{equation}
  \sum_{b,c} f^{abc}f^{dbc} = C_2(\mathrm{adj})\,\delta^{ad},
\end{equation}
and for $\SU{N}$ the quadratic Casimir of the adjoint representation is
$C_2(\mathrm{adj}) = N$, so that
\begin{equation}
  \sum_{b,c} f^{abc}f^{dbc} = N\,\delta^{ad}.
\end{equation}
Setting $d = a$ with no sum on $a$ gives, for each fixed $a$,
$\sum_{b,c=1}^{15} |f^{abc}|^2 = N = 4$ for $\SU{4}$. Splitting this full sum
into light-light, heavy-heavy, and mixed light-heavy index pairs gives the
universal sum rule
\begin{equation}
  C_L^a + C_H^a + C_\mathrm{mix}^a = 4
  \qquad \text{for every generator index } a,
  \label{eq:sumrule_app}
\end{equation}
where $C_\mathrm{mix}^a$ collects the contributions with one index in $\pi_{\rm HC}^V$
and one in $\PiT^H$. For the states considered below the mixed weight
vanishes, and knowledge of $C_L^a$ alone then determines the full decay
pattern.

\subsection*{C.2\quad Evaluating the group weights from the commutation relations}
Since $f^{abc}$ is totally antisymmetric, the commutator
$[\widetilde{\mathcal{X}}^b, \widetilde{\mathcal{X}}^c] \propto
\widetilde{\mathcal{T}}^a$ is nonzero if and only if $f^{abc} \neq 0$. A
nonzero contribution to $C_L^a$ therefore requires that the commutator of two
light-pion generators $K^{V,(b)}$ and $K^{V,(c)}$ produces the generator
$\widetilde{\mathcal{T}}^a$. From the commutation relation in
Table~\ref{tab:broken-generators},
\begin{equation}
  [K^{V,(b)},\, K^{V,(c)}] = \frac{i}{\sqrt{2}}\,\eps^{bca}\,T^{V,(a)},
  \label{eq:KVcomm_app}
\end{equation}
the light-light commutators close only onto the vector isotriplet $T^{V}$,
so only the isotriplet states $\rhoT^{\pm,0}$ receive a light-light
contribution. For fixed $a$ the ordered sum contains exactly two nonzero
terms, so that $\sum_{b,c} |f^{abc}|^2 = \tfrac{1}{2}\sum_{b,c} |\eps^{bca}|^2 = 1$,
giving $C_L = 1$ for the isotriplet
$\rhoT^{\pm,0}$ shown in the first row of Table~\ref{tab:decay_app}. For the same
states, inspection of Table~\ref{tab:commutators_app} shows that the
commutator of one light and one heavy generator never produces the isotriplet
$T^V$, so the mixed weight vanishes, $C_\mathrm{mix} = 0$. The sum
rule~\eqref{eq:sumrule_app} then fixes $C_H = 4 - C_L = 3$.

The same table identifies the heavy channels responsible for $C_H = 3$. The
axial isotriplet generators close onto the vector isotriplet,
$[K^{A,(b)},K^{A,(c)}] = \frac{i}{\sqrt{2}}\,\eps^{bca}\,T^{V,(a)}$, and the same-sector
heavy-pion pairs likewise produce it,
$[K^{I,(i)}, K^{I,(j)}] = \frac{i}{\sqrt{2}}\,\eps_{ijk}\,T^{V,(k)}$ and
$[K^{II,(i)}, K^{II,(j)}] = \frac{i}{\sqrt{2}}\,\eps_{ijk}\,T^{V,(k)}$, each contributing
$\tfrac{1}{2}\sum_{b,c}|\eps^{bca}|^2 = 1$, confirming that
these heavy-heavy channels saturate $C_H = 3$ for $\rhoT^{\pm,0}$.

For the remaining twelve $\rhoT$ states the light-light commutators of
Eq.~\eqref{eq:KVcomm_app} never produce the associated unbroken generator,
since they close only onto $T^V$. Their light-light group weight therefore
vanishes, $C_L^a = 0$, and their decays proceed entirely through channels
involving heavy pions, with weights satisfying
$C_H^a + C_\mathrm{mix}^a = 4$.
This counting is repeated for each unbroken generator in turn, and the resulting
weights $C_L$ and $C_H + C_\mathrm{mix}$ of all fifteen spin-1 resonances, together
with the decay channels that saturate them, are summarized in
Table~\ref{tab:decay_app}.

\begin{table}[!htbp]
\small
\centering
\caption{Decay modes of the fifteen spin-1 resonances, together with their group-theoretic weights. The weight $C_L$ is associated with decays into a pair of light pions $\pi_{\rm HC}^{V}$, identified with $W^\pm$ and $Z^0$, while $C_H$ and $C_\mathrm{mix}$ denote the weights for decays into a pair of heavy pions $\PiT^H$ and into mixed light-heavy pairs, respectively; the two are shown combined in the last weight column. They satisfy the sum rule $C_L + C_H + C_\mathrm{mix} = 4$ of Eq.~\eqref{eq:sumrule_app}. The mixed weight vanishes for the isotriplet $\rhoT^{\pm,0}$ and for the three $\SU{2}_V$ singlets. The first row corresponds to the isotriplet discussed in Section~\ref{sec:branching}, whose branching ratio is given by Eq.~\eqref{eq:BR_final}.}
\label{tab:decay_app}
\renewcommand{\arraystretch}{1.4}
\setlength{\extrarowheight}{3pt}
\setlength{\tabcolsep}{5pt}
\begin{tabular}{l l c c c >{\setlength{\baselineskip}{1.45\baselineskip}}p{7.2cm}}
\hline
\textbf{Generator} & \textbf{State} & $\mathbf{\SU{2}_V}$ & $C_L$ & $C_H + C_\mathrm{mix}$
  & \textbf{Dominant decay channels} \\[3pt]
\hline\hline
$T^{V,(a)}$
  & $\rhoT^{a}$
  & $\mathbf{3}$ & $1$ & $3$
  & $\pi^{V,(a)}_{\rm HC}\pi^{V,(b)}_{\rm HC}$\;$(W^\pm,Z^0)$;\;
    $\PiT^{A,(a)}\PiT^{A,(b)}$;\newline
    $\PiT^{K^{\mathrm{I},(i)}}\PiT^{K^{\mathrm{I},(j)}}$;\;
    $\PiT^{K^{\mathrm{II},(i)}}\PiT^{K^{\mathrm{II},(j)}}$ \\[3pt]
\hline
$T^{A,(a)}$
  & $\hat{\rho}_{\rm HC}^{A,(a)}$
  & $\mathbf{3}$ & $0$ & $4$
  & $\pi^{V,(a)}_{\rm HC}{+}\PiT^{A,(b)}$;\;
    $\PiT^{K^{S,\mathrm{II}}}{+}\PiT^{K^{\mathrm{I},(a)}}$;\newline
    $\PiT^{K^{S,\mathrm{I}}}{+}\PiT^{K^{\mathrm{II},(a)}}$ \\[3pt]
\hline
$X^{\mathrm{I},(i)}$
  & $\hat{\rho}_{\rm HC}^{X^{\mathrm{I},(i)}}$
  & $\mathbf{3}$ & $0$ & $4$
  & $\pi^{V,(a)}_{\rm HC}{+}\PiT^{K^{\mathrm{I},(i)}}$;\;
    $\PiT^{K^{S,\mathrm{II}}}{+}\PiT^{A,(a)}$;\newline
    $\PiT^{K^{S,0}}{+}\PiT^{K^{\mathrm{II},(i)}}$ \\[3pt]
\hline
$S^{\mathrm{I}}$
  & $\hat{\rho}_{\rm HC}^{S^{\mathrm{I}}}$
  & $\mathbf{1}$ & $0$ & $4$
  & $\delta_{ia}\PiT^{K^{\mathrm{II},(i)}}{+}\PiT^{A,(a)}$;\;
    $\PiT^{K^{S,0}}{+}\PiT^{K^{S,\mathrm{II}}}$ \\[3pt]
\hline
$X^{\mathrm{II},(i)}$
  & $\rhoT^{X^{\mathrm{II},(i)}}$
  & $\mathbf{3}$ & $0$ & $4$
  & $\pi^{V,(a)}_{\rm HC}{+}\PiT^{K^{\mathrm{II},(i)}}$;\;
    $\PiT^{K^{S,\mathrm{I}}}{+}\PiT^{A,(a)}$;\newline
    $\PiT^{K^{S,0}}{+}\PiT^{K^{\mathrm{I},(i)}}$ \\[3pt]
\hline
$S^{\mathrm{II}}$
  & $\rhoT^{S^{\mathrm{II}}}$
  & $\mathbf{1}$ & $0$ & $4$
  & $\delta_{ia}\PiT^{K^{\mathrm{I},(i)}}{+}\PiT^{A,(a)}$;\;
    $\PiT^{K^{S,0}}{+}\PiT^{K^{S,\mathrm{I}}}$ \\[3pt]
\hline
$S^{0}$
  & $\hat{\mathcal{X}}_{\rm HC}^{S^{0}}$
  & $\mathbf{1}$ & $0$ & $4$
  & $\PiT^{K^{\mathrm{I},(i)}}{+}\PiT^{K^{\mathrm{II},(i)}}$\;$(\textstyle\sum_i)$;\;
    $\PiT^{K^{S,\mathrm{I}}}{+}\PiT^{K^{S,\mathrm{II}}}$ \\[3pt]
\hline\hline
\multicolumn{6}{p{0.95\textwidth}}{Only $\rhoT^{\pm,0}$ (associated with $T^{V,(a)}$) has $C_L \neq 0$ and couples to $W^\pm Z,\,W^+W^-$.} \\[3pt]
\hline
\end{tabular}
\end{table}

\subsection*{C.3\quad The branching ratio formula}
In the ratio of partial widths formed from Eq.~\eqref{eq:Gamma_app}, all
factors common to every channel cancel, leaving only the group weight
$C_\mathcal{S}^a$ and the $P$-wave phase-space factor $\beta_\mathcal{S}^3$.
For the isotriplet states, with the two channel sets
$\mathcal{S} \in \{\pi_{\rm HC}^V\pi_{\rm HC}^V,\,\PiT^H\PiT^H\}$ and the equal-mass
momentum of Eq.~\eqref{eq:beta_app}, the branching ratio into a set
$\mathcal{S}$ is
\begin{equation}
  \mathrm{BR}(\rhoT^a \to \mathcal{S})
  = \frac{C_\mathcal{S}^a\;\beta_\mathcal{S}^3}
         {\displaystyle\sum_{\mathcal{S}'} C_{\mathcal{S}'}^a\;\beta_{\mathcal{S}'}^3},
  \qquad
  \beta_\mathcal{S} = \sqrt{1 - \frac{4M_\mathcal{S}^2}{\MrhoT^2}},
  \label{eq:BR_app}
\end{equation}
where $\beta_\mathcal{S}$ is the velocity of the final-state pions in channel
set $\mathcal{S}$, with $M_\mathcal{S} = M_W$ for the light channel, in which
$\pi^V_{\rm HC} \sim W^\pm, Z^0$, and $M_\mathcal{S} = M_{\PiT^H}$ for the heavy
channel. The sum in the denominator runs over the kinematically open channel
sets, closure at threshold being enforced automatically by
$\beta_{\mathcal{S}'}^3 \to 0$.

\section{Kinematics}
\label{app:kinematics}

\subsection*{D.1\quad Jet grooming}
\begin{figure}[htbp!]
    \centering
    \begin{subfigure}[b]{0.49\linewidth}
        \centering
        \includegraphics[width=\linewidth]{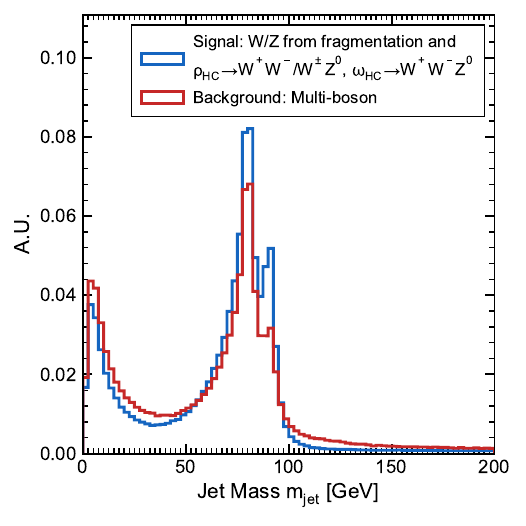}
        \caption{Jet mass distribution.}
        \label{fig:3}
    \end{subfigure}
    \begin{subfigure}[b]{0.49\linewidth}
        \centering
        \includegraphics[width=\linewidth]{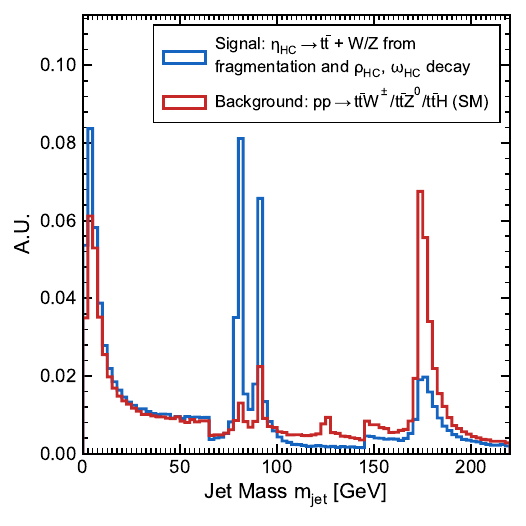}
        \caption{Jet mass distribution with top quark.}
        \label{fig:4}
    \end{subfigure}
    \caption{Groomed jet mass distribution for the signal (blue) against the SM
    background (red): the multi-boson background in (a) and the associated
    top-pair background in (b). }
    \label{FIG:A}
\end{figure}
Each jet is groomed with the SoftDrop algorithm~\cite{Larkoski_2014} using
$z_\mathrm{cut} = 0.1$ and $\beta = 0$, which removes soft wide-angle
radiation and leaves a groomed mass that traces the hard splitting inside the
jet rather than the diffuse spectrum typical of ordinary QCD jets. The
groomed mass distributions in Figure~\ref{FIG:A} exhibit peaks
associated with $W/Z$ jets and top jets, comparing the signal (blue) with the
SM backgrounds (red): the multi-boson and the associated top-pair
background. The signal shows a pronounced peak in the
$W/Z$ mass region, around $80$--$91\,\mathrm{GeV}$, and a second peak near
the top-quark mass of $173\,\mathrm{GeV}$ in events containing the
pseudoscalar resonance, arising from the boosted top quarks produced in
$\etaT \to t\bar{t}$. The SM backgrounds also exhibit pronounced structures in both the EW-boson and top-mass regions. Although the jet-mass distribution identifies the characteristic $W/Z$- and top-jet structures, it does not by itself provide a strong discriminator between signal and background.

\subsection*{D.2\quad Transverse momenta of the decay products}
\begin{figure}[htbp!]
    \centering
    \begin{tabular}{@{}cc@{}}
        \includegraphics[width=0.48\textwidth]{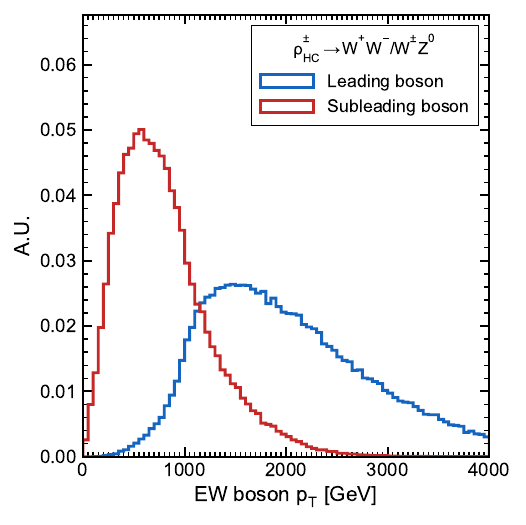} &
        \includegraphics[width=0.48\textwidth]{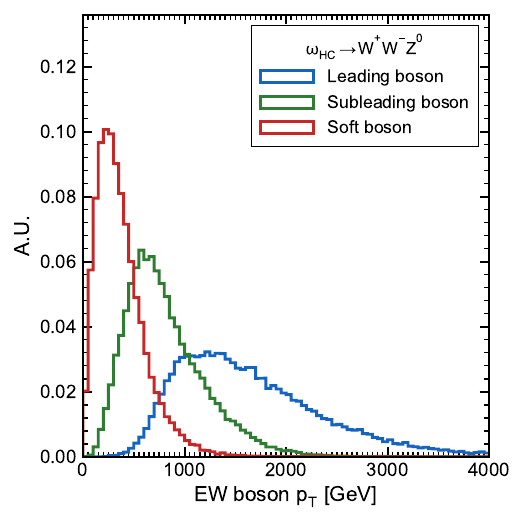}
    \end{tabular}
    \caption{Transverse momentum ($p_T$) distributions of the EW bosons produced in the resonance decays for Benchmark~1, ordered by decreasing $p_T$: the two daughters of the $\rhoT$ (left) and the three daughters of the $\omT$ (right).}
    \label{fig:10}
\end{figure}
Figure~\ref{fig:10} shows the transverse momentum spectra of the resonance
decay products, ordered by decreasing $p_T$. For the isotriplet $\rhoT$, one
daughter carries the larger momentum, of order TeV, approximately $2.5$ times
that of the other. For the isosinglet $\omT$, the momentum is shared among
three daughters, so their $p_T$ spectrum is strongly ordered into a leading, a
subleading and a soft boson. Within $|\eta| \leq 2.5$ the three carry median
$p_T$ of $1557$, $757$ and $343\,\mathrm{GeV}$ at $2\,\mathrm{TeV}$, against
$3157$, $1636$ and $780\,\mathrm{GeV}$ at $5\,\mathrm{TeV}$. The softest boson
therefore limits the reconstruction only at $2\,\mathrm{TeV}$, where its
$16$--$84\%$ range of $168$--$609\,\mathrm{GeV}$ straddles the jet threshold
$p_{T}^{\min} = 250\,\mathrm{GeV}$, while at $5\,\mathrm{TeV}$ it spans
$391$--$1276\,\mathrm{GeV}$; the fraction of decays with all three daughters
resolved as separate jets correspondingly rises from $54\%$ at $2\,\mathrm{TeV}$ to $71\%$ at $5\,\mathrm{TeV}$.

\FloatBarrier
\bibliographystyle{utphys}
\bibliography{biblio}
\end{document}